\documentclass[a4paper,11pt]{article}
\pdfoutput=1
\usepackage{jcappub}
\usepackage[T1]{fontenc}
\usepackage{multirow}
\usepackage{booktabs}
\usepackage{mathtools}
\usepackage{adjustbox}
\usepackage{array}
\usepackage{caption}
\usepackage{slashed}
\usepackage{mathrsfs}

\newcommand{\CLASS}{\texttt{CLASS}}

\newcommand{\hpt}{H_{\rm PT}}

\usepackage{mathtools}
\title{Cosmological and Astrophysical Constraints on Late First-Order Phase Transitions}

\author[a]{Kylar Greene,}
\author[b]{Daven Wei Ren Ho,}
\author[c]{Soubhik Kumar,}
\author[b]{Yuhsin Tsai}

\affiliation[a]{Department of Physics and Astronomy, Seoul National University, 1 Gwanak-ro, Gwanak-gu, Seoul 08826, Korea}
\affiliation[b]{Department of Physics and Astronomy, University of Notre Dame, Notre Dame, IN 46556, USA}
\affiliation[c]{Institute of Cosmology, Department of Physics and Astronomy, Tufts University, Medford, MA 02155, USA}

\begin{document}

\abstract{First-order cosmological phase transitions (PT) can take place in a dark sector at relatively late times between the big-bang nucleosynthesis and recombination epochs. Because bubble nucleation is stochastic, the PT completes at different times in different regions of the Universe. This fluctuation sources a curvature perturbation whose (dimensionless) power spectrum ${\cal P}_\zeta(k)$ features a universal infrared tail, independent of the microscopic details of the PT. 
Even in the absence of any non-gravitational interaction between the dark sector and the Standard Model, these additional curvature perturbations at small scales impact a variety of observables.
We derive new constraints on dark sector phase transitions using {\it Planck}, baryon acoustic oscillation (BAO), Lyman-$\alpha$ observations, spectral distortion limits from FIRAS, constraints on early reionization, and the existence of ultra-faint dwarf galaxies.}

\maketitle

\section{Introduction}
First-order phase transitions (PT) in the early Universe have long been studied for a variety of reasons---from the possibility that electroweak symmetry breaking proceeds via a first-order transition in extensions of the Standard Model \cite{Kirzhnits:1972ut, Dolan:1973qd, Quiros:1999jp, Weir:2017wfa, Gould:2019qek, Gould:2021oba, Gould:2022ran, Camargo-Molina:2021zgz, Qin:2024dfp}, to their potential to explain the baryon asymmetry of the Universe \cite{Kuzmin:1985mm, Cohen:1993nk, Morrissey:2012db, Damgaard:2015con, deVries:2017ncy, Cline:2017qpe, Cline:2017jvp, Ellis:2019flb, Croon:2019ugf, Fuchs:2020uoc, Bodeker:2020ghk, Wang:2022ygk, Bittar:2024nrn, Gao:2024fhm, Biswas:2025rzs, vandeVis:2025efm}, and to their ability to generate stochastic gravitational wave (GW) backgrounds observable by current \cite{EPTA:2021crs, Reardon:2023gzh, NANOGrav:2023gor, Xu:2023wog, LIGOScientific:2025bgj} and future detectors \cite{Sathyaprakash:2011bh, Janssen:2014dka, Caprini:2015zlo, Caprini:2019egz, Reitze:2019iox, Kawamura:2020pcg, SKAOPulsarScienceWorkingGroup:2025oyu}. The majority of these studies focus on transitions occurring at temperatures $\gtrsim$ MeV.

Motivated by emerging tensions in cosmological measurements~\cite{Riess:2021jrx,Schoneberg:2021qvd,Niedermann:2019olb,Niedermann:2020dwg,Niedermann:2023ssr,Garny:2024ums,Chatrchyan:2024xjj} and by the search for post-inflationary physics capable of generating both CMB $B$-modes and scalar perturbations \cite{Pogosian:2007gi,Pogosian:2018vfr,Domenech:2021ztg,Silva:2023diq,Ireland:2025yqr,Greene:2026one}, there has been growing interest in \emph{late-time} PTs occurring at temperatures $\lesssim$ MeV, corresponding to redshifts $z_{\rm PT} \lesssim 10^9$. Such late-time transitions must take place entirely within a dark sector to avoid excessive reheating of SM particles \cite{Poulin:2016anj,Breitbach:2018ddu,Ertas:2021xeh,Bai:2021ibt,Hill:2023wda}. As a result, conventional detection methods, such as collider experiments, that rely on direct SM couplings are ineffective for probing these secluded scenarios. The dark sector still couples gravitationally to the SM, modifying photon and baryon density perturbations, which makes astrophysical and cosmological observations powerful probes of possible dark sector PTs. While GW observatories, such as the Laser Interferometer Space Antenna (LISA), probe the PT dynamics directly, the signatures due to enhanced curvature perturbations that we study provides complementary sensitivity that extends to weaker PTs.
In these cases, the GW signature itself falls below detection thresholds, but the cosmological and astrophysical signatures we discuss can be observable. 

A first-order PT proceeds via nucleation of bubbles which subsequently percolate, leading to a completion of the PT.
However, bubble nucleation is a stochastic process, and in different causally disconnected regions of the Universe, the PT does not complete exactly at the same time.
It was shown in~\cite{Liu:2022lvz,Elor:2023xbz,Lewicki:2024ghw,Franciolini:2025ztf} that these fluctuations in the PT completion time source curvature perturbations, and for late enough PTs, there would be distinct imprints on the cosmic microwave background (CMB) and large-scale structure~\cite{Elor:2023xbz,Koren:2025ymq}.
This allowed using existing measurements to place stringent constraints on the energy released during a PT. 
As we review below, the power spectrum of these curvature perturbations is peaked at the scale of the average bubble separation and exhibits a well-defined tail ($k^3$, if the dark sector dilutes as radiation after the PT) on larger scales. 
This characteristic scale dependence can distinguish it from perturbations generated during inflation, making it a well-motivated scenario for post-inflationary production of curvature perturbations that are uncorrelated with the adiabatic modes.

In this work, we refine and extend the constraints derived in~\cite{Elor:2023xbz} along several directions. After reviewing how a PT sources curvature perturbations, we derive bounds on the resulting perturbations from a variety of astrophysical and cosmological observations. We obtain CMB and BAO constraints on late-time dark sector PTs by performing a Markov Chain Monte Carlo (MCMC) likelihood analysis, varying $\Lambda$CDM parameters alongside the PT parameters to properly take into account parameter degeneracies.
For Lyman-$\alpha$ bounds, in addition to adopting existing curvature-perturbation constraints from~\cite{2011MNRAS.413.1717B} which parameterize the primordial power spectrum as a piecewise-linear function, we carry out a dedicated two-parameter fit of the PT plus SM contribution to the compressed eBOSS flux power spectrum~\cite{Bird:2023,Fernandez:2024,He:2025}.
Beyond the FIRAS limits on CMB spectral distortions studied in~\cite{Elor:2023xbz}, we also derive new constraints by requiring that the PT-generated density fluctuations do not induce too early reionization~\cite{Qin:2025ymc} or excessively evaporate observed ultra-faint dwarf (UFD) galaxies through enhanced small-scale perturbations~\cite{Graham:2023unf,Graham:2024hah}. 
Together, these improvements extend existing exclusion bounds to include scenarios with even smaller amounts of energy release, thus weaker PTs, and PTs that take place at even earlier times.
Future cosmological probes will further improve the sensitivity to late-time PTs.

The paper is organized as follows. In Section~\ref{sec:darkFOPT}, we review how the time fluctuations in the completion of a first-order PT generate isocurvature perturbations that then source curvature perturbations. Utilizing the power spectrum of these perturbations, in Section~\ref{sec.analysis} we summarize our derived constraints on a dark sector PT. We then detail the derivation of the constraints from CMB in Section~\ref{sec:CMB}, including an MCMC analysis of combined CMB and BAO data, as well as estimates from CMB spectral distortions. In Section~\ref{sec:Struct}, we discuss constraints from structure formation, including Lyman-$\alpha$ bounds, limits from early reionization, and constraints derived from observations of ultra-faint dwarf galaxies. We conclude in Section~\ref{sec:conclusion}.

\section{Curvature perturbation from a first-order phase transition}
\label{sec:darkFOPT}
We focus on scenarios where a secluded dark sector, observable only through its gravitational effects, is initially in a false vacuum and undergoes a PT during the radiation-dominated era. The nucleation rate per volume can be described by~\cite{HOGAN1983172, PhysRevD.45.3415,Hindmarsh:2015qta}:  
\begin{equation}\label{eq.Gamma}
\Gamma = \Gamma_0 e^{-S(t)} \approx \Gamma_0 e^{-S(t_f)}e^{\beta(t-t_f)}
\end{equation}
where $S$ is the bounce action for nucleating a bubble. The reference time scale $t_f$ is conventionally taken as the time when a $1/e$ fraction of space remains in the false vacuum~\cite{ATHRON2024104094}, and is approximately the Hubble time of the phase transition, $\hpt \equiv H(t_f)$ where $H(t)$ is the Hubble scale at time $t$. The PT completion rate is quantified by $\beta \approx -{\rm d} S(t_f)/{\rm d} t$ such that when $\beta \gg \hpt$, the phase transition proceeds rapidly and completes within a small fraction of the Hubble time near $t_f$. Correspondingly, $1/\beta$ sets the PT duration time scale and the average size reached by the expanding bubbles within that time goes as $d_b \approx (8\pi)^{1/3}v_w/\beta$, where $v_w$ is the bubble wall velocity~\cite{PhysRevD.45.3415,Hindmarsh:2015qta}. 

Since bubbles are nucleated stochastically, different regions of space generally transition at different times as the PT unfolds. 
We can use this to study the statistics of bubble nucleation before considering the energetics of this process. 
Let us define the transition time at a spatial point $\vec{x}$ as $t_c(\vec{x}) =\bar{t}_c + \delta t_c(\vec{x})$, where $\bar{t}_c$ is the space averaged transition time and $\delta t_c(\vec{x})$ is the fluctuation around it. We assume these fluctuations are small compared to the PT time such that $\delta t_c(\vec{x}) \ll \bar{t}_c$.
The statistics of these fluctuations can be quantified using the two-point correlator $\hpt^2 \langle \delta t_c(\vec{x}) \delta t_c(\vec{y}) \rangle$, with the corresponding dimensionless power spectrum~\cite{Elor:2023xbz}:
\begin{equation}\label{eq.Pdt}
\mathcal{P}_{\delta t}(k) = \frac{k^3}{2\pi^2} \left(\frac{\hpt}{\beta}\right)^2 \int \mathrm{d}^3 r~ e^{i \vec{k} \cdot \vec{r}} \beta^2 \langle \delta t_c(\vec{x}) \delta t_c(\vec{y}) \rangle
\end{equation}
where $\vec{r} = \vec{x} - \vec{y}$ and $k$ is the comoving wavenumber. Wavelengths much larger than the comoving bubble size at PT completion $k^{-1} \gg d_b/a_{\rm PT}$, where $a_{\rm PT} = a(t_f)$, would span a large but finite number of bubbles. For perturbation modes on these scales, the power spectrum would then reflect the statistical variance in PT completion time, which goes as $\langle \delta t_c^2 \rangle\sim 1/N$ for $N$ uncorrelated transitions in each volume of size $\sim 1/k^3$. This leads to a universal scaling~\cite{Elor:2023xbz}:  
\begin{equation}\label{eq:Pdt}  
{\cal P}_{\delta t}(k) \approx 3(8\pi) v_w^3 \left(\frac{k}{a_{\rm PT} \hpt}\right)^3 \left(\frac{\hpt}{\beta}\right)^5  
\end{equation}
on super-bubble scales. Notably, the amplitude depends explicitly on the PT completion rate via the dimensionless ratio $\beta/H_{\rm PT}$, since this directly affects the average size of bubbles at PT completion and hence the number of bubbles in each comoving volume. For wavelengths much shorter than the average bubble size $k^{-1}\ll d_b/a_{\rm PT}$, the power spectrum was found to be proportional to ${\cal P}_{\delta t}(k)\propto (k/(a_{\rm PT} \hpt))^{-3}(H_{\rm PT}/\beta)^{-1}$ instead~\cite{Elor:2023xbz}. More generally, the light-cone geometry of the expanding bubbles can be analyzed to obtain an analytic expression for the $\delta t_c$ correlator~\cite{Elor:2023xbz,Jinno:2017,Jinno:2019}, which can then be integrated numerically to obtain the full ${\cal P}_{\delta t}(k)$ spectrum\footnote{Further details on the numerical calculation can be found in Appendix~\ref{app:numerical}}. In the following analysis, we use the numerically integrated power spectrum obtained in~\cite{Elor:2023xbz}.

Let us now account for the energy content of the dark sector over the course of the PT. Prior to the PT, the dark sector is in a false vacuum which, by definition, is at a higher potential than the true vacuum. This potential difference provides a false vacuum energy $F$ that redshifts as a cosmological constant $\rho_F \sim \rm const$, and which is then liberated as the latent heat of the PT when the dark sector transitions to the true vacuum. The PT strength is quantified by the ratio of the latent heat to the remainder of the energy present in the universe at the PT:  
\begin{equation}
    \alpha_{\rm PT} \equiv \frac{\rho_F}{\rho_{\rm SM, PT}}
\end{equation}
where $\rho_{\rm SM, PT}=\rho_{\rm SM}(t_f)$ is the standard model radiation at PT completion, which makes up the dominant part of the energy in the radiation era. For simplicity, we assume that the dark sector is dominated by $F$ initially, and after the PT the latent heat goes entirely into populating it with some dark fluid component $X$ that redshifts generally as $\rho_X \sim a^{-4-d_X}$, for $d_X$ the deviation from radiation-like behavior. Requiring that the energy density is continuous at the transition, the energy density in this dark sector is described by
\begin{align}
\label{eq:rhoX}
    \rho_X(\vec{x}) = 
    \begin{cases}
        \rho_F \hspace{40mm}(t<t_c(\vec{x}))\\
        \rho_F\left( \frac{a(t)}{a_c(\vec{x})} \right)^{-(4+d_X)} \hspace{14mm} (t\geq t_c(\vec{x}))
    \end{cases}
\end{align}
where we denote $a_c(\vec{x}) \equiv a(t_c(\vec{x}))$ and $\bar{a}_c \equiv a(\bar{t}_c)$ for the background scale factor. Combining this with the fact that the PT time $t_c(\vec{x})$ fluctuates across space, the energy content of the dark sector likewise transitions from the false vacuum $F$ to the dark fluid $X$ at different times across space. This generates density perturbations $\delta\rho_X(\vec{x})$ in the dark fluid $X$ by the PT completion time $t_f$.
\begin{figure}
    \centering
    \includegraphics[width=0.8\linewidth]{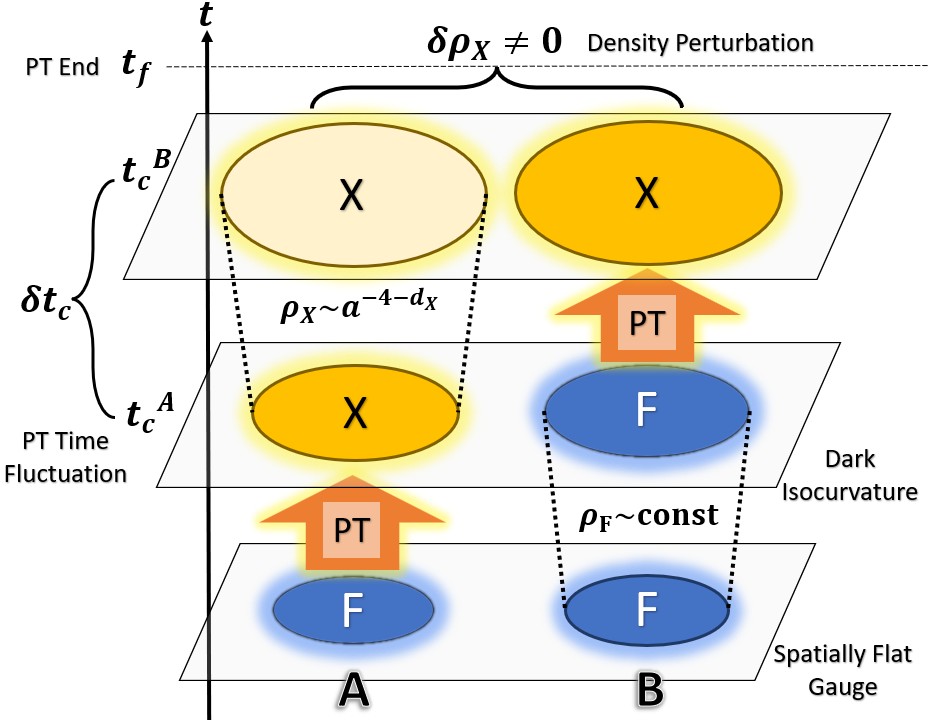}
    \caption{Schematic picture of dark first-order PT in two far apart patches, $A$ and $B$, initially in the false vacuum $F$. The transition from $F$ to dark fluid $X$ occurs in each patch at times $t^A_c$ and $t^B_c$, respectively. The PT generates an isocurvature perturbation between the two patches at $t^A_c$. This isocurvature perturbation sources a curvature (density) perturbation between $t^A_c$ and $t^B_c$, whenever $F$ and $X$ redshift differently.}
    \label{fig:cartoon}
\end{figure}

An intuitive picture of how this process unfolds on large scales can be obtained via the separate universe approach~\cite{Elor:2023xbz}. For this argument, we ignore the standard curvature perturbations generated during inflation.
In spatially flat gauge, let us consider sufficiently separated patches $A$ and $B$ that undergo PT at different times, as shown schematically in Fig.~\ref{fig:cartoon}. Suppose that at a time $t^A_c$, the vacuum $F$ transitions into the dark fluid $X$ in patch $A$, but patch $B$ has not transitioned yet and remains in $F$ until a later time $t^B_c$. This means that at $t^A_c$, while there is no curvature perturbation, there is an isocurvature perturbation since the composition of the two patches are different. During the time interval $t^A_c < t < t^B_c$, the dark sector thus consists of a mixture of components, $F$ and $X$, that redshift differently as a function of time.
This induces a relative difference in energy densities between the patches and sources a curvature perturbation by the time PT completes at $t_f$. To compute this, note the energy densities in the two patches are given by
\begin{align}
\label{eq:rho_patch}
    \rho^{i}_X(t_f) = \rho_F \left(\frac{t^{i}_c}{t_f}\right)^{2+d_X/2} \quad\quad\quad t^{i}_c \leq t_f,\quad i \in \{A, B\}
\end{align}
where $t\sim a^2$ in the radiation era. Notably, we see that $t^B_c\neq t^A_c$ implies $\rho^{A}_X(t_f) \neq \rho^{B}_X(t_f)$, so a fluctuation in PT time between the patches ultimately sources a density perturbation in the dark fluid $X$ after PT completes. Generalizing this from discrete patches $i$ to the continuous variable $\vec{x}$, the density perturbation $\delta\rho_X(\vec{x})$ sourced by the PT can be obtained by varying with respect to $t_c(\vec{x})$ in Eq.~(\ref{eq:rhoX}):
\begin{align}
\label{eq:rho_delta}
    \frac{\delta\rho_X(\vec{x})}{\rho_X} \approx (4+d_X) H_{\rm PT} \delta t_c(\vec{x})\,,
\end{align}
assuming short PT duration $\delta t_c \ll \bar{t}_c$, and using $H_{\rm PT}\approx 1/2\bar{t}_c$. 

For a secluded dark sector that interacts with us only gravitationally, we do not observe the fluctuations in the fluid $X$ directly. Instead, we consider the effects that the dark PT has on cosmological observables through the scalar curvature perturbations that it generates on large scales at the PT time. While we gave a heuristic `separate universe' argument above, we now adopt a gauge-invariant description. The curvature perturbation on uniform-density hypersurfaces $\zeta(\vec{x})$, is sourced on large (super-horizon) scales according to the equation~\cite{Wands:2000dp}:
\begin{align}
\label{eq:wands}
    \dot{\zeta}(\vec{x}) = - \frac{H}{\rho + p}\delta p_{\rm nad}(\vec{x})
\end{align}
where
\begin{align}
    \delta p_{\rm nad}(\vec{x}) = \delta p(\vec{x}) - \frac{\dot{p}}{\dot{\rho}}\delta\rho(\vec{x})
\end{align}
is the non-adiabatic component of the total pressure perturbation, with dot denoting derivative with respect to physical time $t$. $\delta p_{\rm nad}$ is the isocurvature perturbation of the full cosmic fluid, and is non-vanishing when the equation of state $\omega \equiv p/\rho$ of the fluid is able to change independently of its density $\rho$. As we have seen, a first-order PT (FOPT) introduces $\delta p_{\rm nad}$ whenever $F$ and $X$ redshift differently and $\delta t_c(\vec{x})$, being stochastic, is uncorrelated to any other perturbations. 

Let us denote $\zeta_{\rm PT}(\vec{x})$ with the subscript ``PT'' to distinguish the component of the curvature perturbation sourced specifically from the dark PT; we will add the $\zeta$ from the standard inflationary dynamics later.
To evaluate $\zeta_{\rm PT}(\vec{x})$ explicitly, we can use Eq.~(\ref{eq:rhoX}) and $\rho = \rho_{\rm SM} + \rho_X$ to write:
\begin{align}
    \rho(\vec{x}) =&\ \rho_F \Theta(t_c(\vec{x}) - t) + \rho_{F}\left( \frac{a(t)}{a_c(\vec{x})} \right)^{-(4+d_X)} \Theta(t - t_c(\vec{x})) + \rho_{\rm SM, PT} \left( \frac{a(t)}{a_{\rm PT}} \right)^{-4}\\
    p(\vec{x}) =& -\rho_F \Theta(t_c(\vec{x}) - t) + \left(\frac{1}{3} + \frac{d_X}{3}\right)\rho_{F}\left( \frac{a(t)}{a_c(\vec{x})} \right)^{-(4+d_X)} \Theta(t - t_c(\vec{x})) + \frac{1}{3}\rho_{\rm SM, PT} \left( \frac{a(t)}{a_{\rm PT}} \right)^{-4}
\end{align}
where $p = \sum_i\omega_i\rho_i$ for the pressure with $\omega_{\rm SM} = 1/3$ for radiation, $\omega_F = -1$ for the false vacuum energy, and $\omega_X = (1+d_X)/3$ for the dark fluid $X$. The density and pressure perturbations can be derived by taking linear variations with respect to $t_c(\vec{x})$, as we had done for Eq.~(\ref{eq:rho_delta}). Leaving further details for the calculation to Appendix~\ref{app:Pnad}, we can plug in the expressions for the source term in Eq.~(\ref{eq:wands}) to obtain:
\begin{align}
\label{eq:zeta_dot}
    \dot{\zeta}_{\rm PT}(\vec{x}) = \left(\frac{\frac{4+d_X}{4}\alpha_{\rm PT}\delta(t-\bar{t}_c)}{\left(1+\frac{4+d_X}{4}\alpha_{\rm PT}\Theta(t-\bar{t}_c)\right)^2}\ +\frac{-Hd_X\frac{4+d_X}{4}\alpha_{\rm PT}\left(\frac{a(t)}{a_{\rm PT}}\right)^{-d_X}\Theta(t-\bar{t}_c)}{\left(1+\frac{4+d_X}{4}\alpha_{\rm PT}\left(\frac{a(t)}{a_{\rm PT}}\right)^{-d_X}\Theta(t-\bar{t}_c)\right)^2}\right)\frac{\delta t_c(\vec{x})}{2\bar{t}_c}
\end{align}
for $\bar{t}_c\approx t_f$ and $a_c \approx a_{\rm PT}$. Notably, the source term has two parts corresponding to the ways in which $\omega$ can change in a way uncorrelated to the adiabatic fluctuations: (i) a $\delta$-function source for the abrupt change from $F$ to $X$ during the random $\delta t_c(\vec{x})$ fluctuations of the FOPT, and (ii) a step-function source that appears only if the resulting $\delta t_c$-perturbed fluid $X$ subsequently redshifts differently from the rest of the radiation bath $d_X \neq 0$.

For our subsequent phenomenological study, we are interested in the regime where $\alpha_{\rm PT} \ll 1$, since regions of parameter space where the PT is too strong would already be excluded by existing data. To leading order in $\alpha_{\rm PT}$, the curvature perturbation at $t>t_f$ is obtained by integrating over the delta and step functions:
\begin{align}
\label{eq:zeta}
    {\zeta}_{\rm PT}(\vec{x}) &= \frac{4+d_X}{4}\alpha_{\rm PT} \left(1 + \int_{t_f}^t dt'\frac{d}{dt'}\left(\frac{a(t')}{a_{\rm PT}}\right)^{-d_X}\right)\frac{\delta t_c(\vec{x})}{2\bar{t}_c}
    = \alpha_{\rm PT} F_{d_X}(a) H_{\rm PT}\delta t_c(\vec{x}),
\end{align}
where we defined the redshift factor
\begin{equation}
    F_{d_X}(a) \equiv  \frac{4+d_X}{4} \left(\frac{a}{a_{\rm PT}}\right)^{-d_X}.
\end{equation}
Given relation \eqref{eq:zeta} between ${\zeta}_{\rm PT}(\vec{x})$ and $\delta t_c(\vec{x})$, the predicted curvature power spectrum from the dark PT can be written in terms of the ${\cal P}_{\delta t}(k)$ spectrum by (using $\vec{r}=\vec{x}-\vec{y}$):
\begin{align}
\label{eq:Pzeta}
    {\cal P}_{\zeta, \rm PT}(k) &= \frac{k^3}{2\pi^2} \int \mathrm{d}^3 r~ e^{i \vec{k} \cdot \vec{r}} \langle \zeta_{\rm PT}(\vec{x}) \zeta_{\rm PT}(\vec{y}) \rangle= \alpha_{\rm PT}^2\, F_{d_X}(a)^2\, {\cal P}_{\delta t}(k).
\end{align}
We stress that since Eq.~\eqref{eq:Pzeta} was derived assuming Eq.~\eqref{eq:wands}, valid on super-horizon scales, and by ignoring hydrodynamic effects during the PT, we do not expect this to remain an accurate description of the curvature spectrum on scales that are small compared to the horizon at PT, and in particular on sub-bubble scales. 
At these scales, the pressure gradients, bubble wall dynamics, and possible sound waves in the dark fluid will likely affect the curvature spectrum.
One should thus exercise caution when using the spectrum for perturbation modes on scales $k \gtrsim a_{\rm PT}/d_b$. We discuss this in more detail in Sec.~\ref{sec.analysis}.

\begin{figure}
    \centering
    \includegraphics[width=0.75\linewidth]{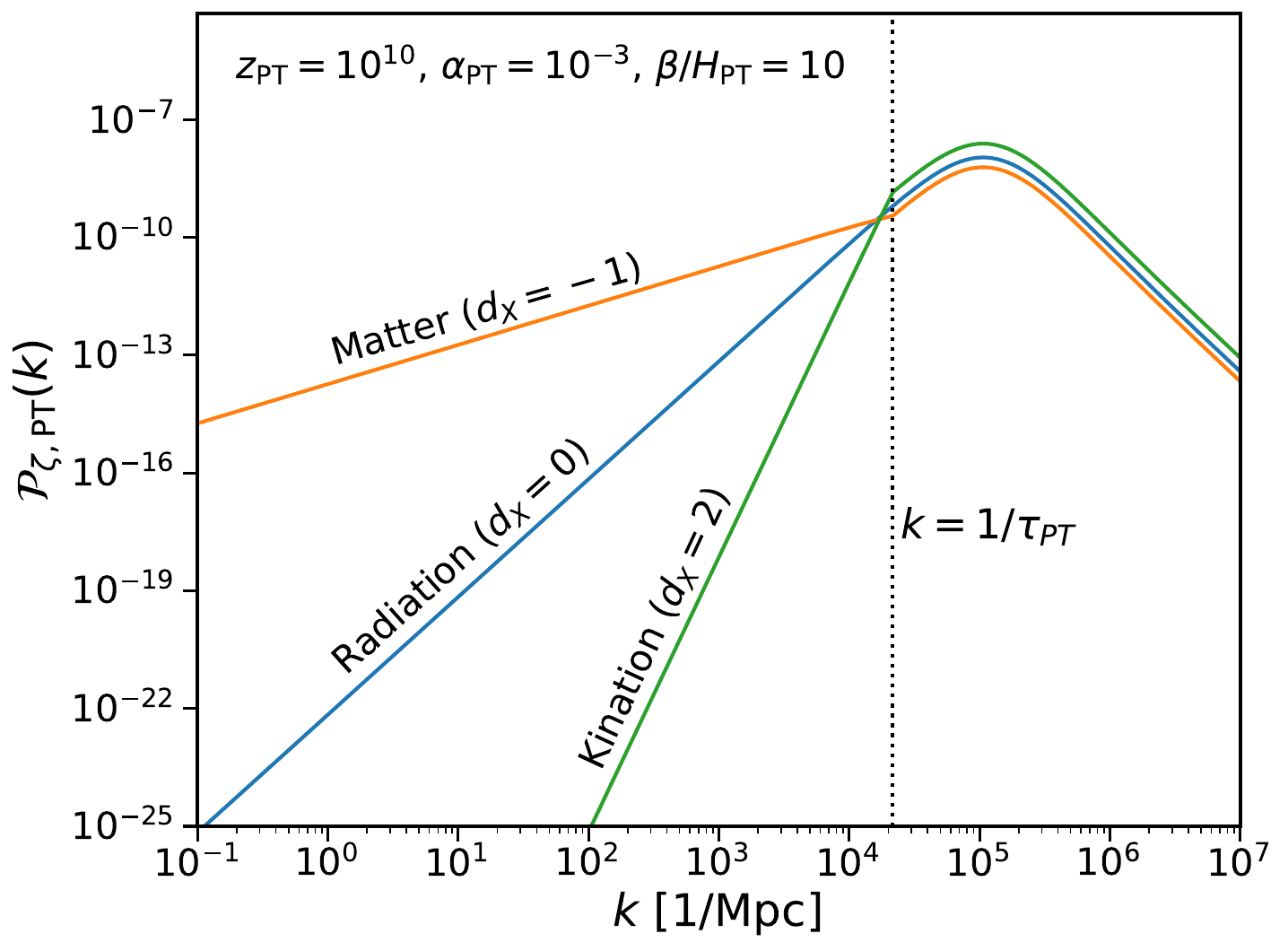}
    \caption{Plot of ${\cal P}_{\zeta, \rm PT}(k)$ for a choice of PT parameters showing modification of super-horizon slope for different redshift behaviors of the dark fluid $X$. We show the cases where $X$ redshifts as radiation ($d_X = 0$), as matter ($d_X = -1$) and as kination ($d_X = 2$). Compared to the $\sim k^3$ power law for radiation, matter exhibits an enhanced $\sim k$ power law in the IR slope while kination exhibits a diminished $\sim k^7$ power law. Slight modifications to the amplitude come from the ${(4+d_X)}/{4}$ prefactor (in Eq.~\eqref{eq:fdx}) due to different redshifts relative to the false vacuum during the $\delta t_c$-fluctuations of the PT. Note that the enhanced IR slope for matter is not expected to hold past matter-radiation equality $k_{\rm eq} \approx 0.01/$Mpc.}
    \label{fig:FdX}
\end{figure}

After the curvature power spectrum ${\cal P}_{\zeta, \rm PT}(k)$ has been generated by the dark PT, perturbation modes $k$ entering the horizon at $\tau \sim 1/k >\tau_{\rm PT}$ will probe the super-horizon region of ${\cal P}_{\zeta, \rm PT}(k)$, where  $\tau \propto a$ and $\tau_{\rm PT} =(a_{\rm PT} H_{\rm PT})^{-1}$ in the radiation era. If the dark fluid $X$ is a dark radiation (DR) component that redshifts the same way as the background SM radiation bath, we have $\omega_{\rm DR} = \omega_{\rm SM} = 1/3$ and $d_X = 0$. The redshift factor in this case is trivial $F_0(a) = 1$ and Eq.~\eqref{eq:Pzeta} implies that the super-horizon slope of the curvature spectrum would just be the $k^3$ slope of ${\cal P}_{\delta t}(k)$ as found in Eq.~(\ref{eq:Pdt}). 
If the dark fluid $X$ has a different redshift factor, this slope can be further modified on super-horizon scales.
We can compute this by evaluating $F_{d_X}$ at horizon entry using $a/a_{\rm PT} = \tau/\tau_{\rm PT} = 1/(k\tau_{\rm PT}),$ giving
\begin{equation}\label{eq:fdx}
    F_{d_X}(k) = \frac{4+d_X}{4} (k \tau_{\rm PT})^{d_X} \quad\quad (k < 1/\tau_{\rm PT}).
\end{equation}
In Fig. \ref{fig:FdX}, we show the modified slopes for the cases where $X$ redshifts as non-relativistic matter $d_X = -1$ and as kination $d_X = 2$ immediately after the PT completes. Since matter redshifts slower than SM radiation ($d_X < 0$), perturbations sourced in matter by the PT make up an increasing contribution to the metric perturbation as time progresses in the radiation era. 
Super-horizon curvature perturbations that enter the horizon at later times are thus larger compared to the scenario of $d_X=0$.
This manifests itself as a gentler $k$ slope in the power spectrum when compared to the $k^3$ slope for radiation-like dilution. The opposite happens for kination in which case the dark sector redshifts faster than SM radiation ($d_X > 0$), resulting in a steeper $k^7$ slope. 
Further, since the different redshift behaviors also change how $X$ redshifts relative to the false vacuum $F$ during the $\delta t_c$-fluctuations of the PT, the strength of the curvature perturbation that is sourced from the isocurvature fluctuations between $F$ and $X$ will also be affected. This is captured by the ${(4+d_X)}/{4}$ prefactor in $F_{d_X}(k)$ which modifies the amplitude of ${\cal P}_{\zeta, \rm PT}(k)$\footnote{As a consistency check, note that if $X=F$, then $d_X = -4$ for cosmological constant and ${\cal P}_{\zeta, \rm PT}(k)$ vanishes since there is no PT.}.   

\section{Summary of observational constraints}\label{sec.analysis}
\begin{figure}
    \centering
    \includegraphics[width=0.8\linewidth]{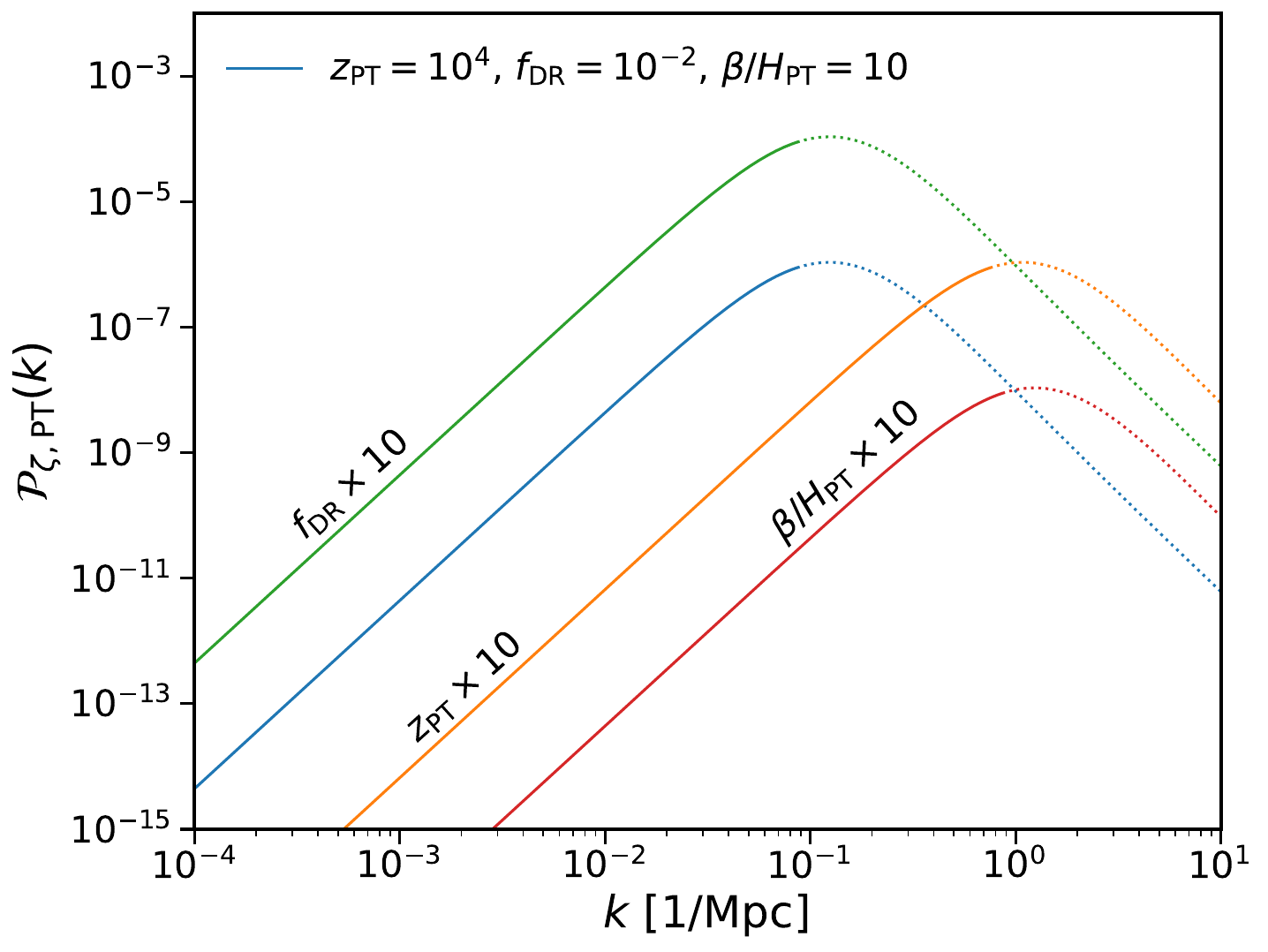}
    \caption{Demonstration of how the curvature power spectrum ${\cal P}_{\zeta, \rm PT}(k)$, for $X={\rm DR}$, changes as the PT parameters $(z_{\rm PT},\ f_{\rm DR},\ \beta/H_{\rm PT})$ are varied. The sub-bubble region of each spectrum is left dotted to stress the uncertainty in ${\cal P}_{\zeta, \rm PT}(k)$ on scales $k \gtrsim a_{\rm PT}/d_b$. All the peaks exhibit the same shape with $k^3$ super-bubble slope. The variations are exaggerated by increasing each of the three PT parameters by an order of magnitude. Using the blue curve as a starting reference, observe that (i) increasing $z_{\rm PT}$ translates the peak position to higher $k$ (yellow), (ii) increasing $f_{\rm DR}$ increases the overall amplitude (green), and (iii) increasing $\beta/H_{\rm PT}$ both translates the peak to higher $k$ and decreases its amplitude (red).}
    \label{fig:Pk_spec}
\end{figure}

In this section we present our main constraints on dark sector PTs, restricting ourselves to the simplest case where the dark fluid $X$ is non-interacting DR that is free-streaming after the PT. 
To study the parameter space of such models, it is convenient to re-parameterize the PT strength in terms of the energy density ratio 
\begin{equation}
    f_{\rm DR} \equiv \frac{\alpha_{\rm PT}}{1+\alpha_{\rm PT}} = \frac{\rho_{\rm DR}}{\rho_{\rm DR} + \rho_\nu + \rho_\gamma}\,,
\end{equation}
which is a constant since the DR, neutrino ($\nu$) and photon ($\gamma$) energy densities on the RHS redshift the same way for the temperatures that we consider. $f_{\rm DR}$ takes values between $0$ and $1$, and is equivalent to the latent heat of the PT normalized over the \textit{total} energy at PT time, as defined in \cite{Elor:2023xbz}. Since $F_0(a) = 1$ for DR, the curvature power spectrum \eqref{eq:Pzeta} is approximately
\begin{equation}
    {\cal P}_{\zeta, \rm PT}(k) = (1+\alpha_{\rm PT})^2\, f_{\rm DR}^2\, {\cal P}_{\delta t}(k)\, \approx f_{\rm DR}^2\, {\cal P}_{\delta t}(k)\,
\end{equation}
to leading order in $\alpha_{\rm PT} \ll 1$. Given the shape of ${\cal P}_{\delta t}(k)$, which can be calculated numerically following Appendix~\ref{app:numerical}, the curvature spectrum ${\cal P}_{\zeta, \rm PT}(k)$ generated by the dark PT can be described \textit{uniquely} with three parameters: 
\begin{itemize}
    \item {\bf $z_{\rm PT}$}: Redshift of the PT, directly related to the SM photon temperature $T_{\rm PT}$. 
    \item {\bf $\beta/H_{\rm PT}$}: Phase-transition rate normalized to the Hubble expansion rate at the transition.
    \item {\bf $f_{\rm DR}$}: Ratio of the energy density released in the PT to that of the total radiation energy density.
\end{itemize}
In Fig.~\ref{fig:Pk_spec}, we plot ${\cal P}_{\zeta, \rm PT}(k)$ to show how the spectrum scales with these parameters. 
The PT signal ${\cal P}_{\zeta, \rm PT}(k)$ is implemented by approximating it as a contribution to the \textit{primordial} curvature power spectrum:
\begin{align}\label{eq.totalP}
    {\cal P}_{\zeta}(k) &= {\cal P}_{\zeta, \rm ad}(k) + {\cal P}_{\zeta, \rm PT}(k)\,,\notag \\
    &= {\cal P}_{\zeta, \rm ad}(k) + f_{\rm DR}^2\frac{k^3}{2\pi^2} \left(\frac{H_{\rm PT}}{\beta}\right)^2 \int \mathrm{d}^3 r~ e^{i \vec{k} \cdot \vec{r}} \beta^2 \langle \delta t_c(\vec{x}) \delta t_c(\vec{y}) \rangle\,,
\end{align}
where ${\cal P}_{\zeta, \rm ad}(k) = A_s(k/k_{\rm pivot})^{n_s -1}$ is the $\Lambda$CDM primordial spectrum for the uncorrelated adiabatic perturbation $\zeta_{\rm ad}$. For calculations where we fix the $\Lambda$CDM parameters, we use the pivot scale $k_{\rm pivot} = 0.05\ \mathrm{Mpc}^{-1}$, scalar amplitude $A_s = 2.1\times10^{-9}$, and spectral index $n_s = 0.966$~\cite{Planck:2018vyg}. We note that by including the PT signal in this way, we are treating it as an \textit{initial} (superhorizon) curvature perturbation with respect to the evolution of cosmological observables. While this is okay for modes with $k\leq1/\tau_{\rm PT}$, it implies we need to include corrections due to sub-horizon mode evolution for modes with $k>1/\tau_{\rm PT}$. Without these corrections, given an initial superhorizon value, such a mode would re-enter the horizon {\it prior} to the PT and start evolving following standard transfer functions. However, in reality, such a mode is not even generated prior to the PT. Thus, to compute the observational effects, we need to `undo' the spurious time evolution, from the time of re-entry to $\tau_{\rm PT}$, that happens when treating ${\cal P}_{\zeta, \rm PT}(k)$ as an initial curvature perturbation. For CMB constraints, since photon perturbation modes that re-enter the horizon earlier suffer significant Silk damping, not applying this correction, which is what we do for simplicity, underestimates the PT signal in the CMB. This makes the CMB bounds studied in Section~\ref{sec:CMB} more conservative. On the other hand, for structure formation constraints, matter perturbations grow (spuriously) after re-entering the horizon. Hence, not applying this correction would overestimate the PT signal. Therefore, we include correction factors to undo this growth for the structure formation bounds studied in Section~\ref{sec:Struct}.

\begin{figure}
    \centering
    \includegraphics[width=0.695\linewidth]{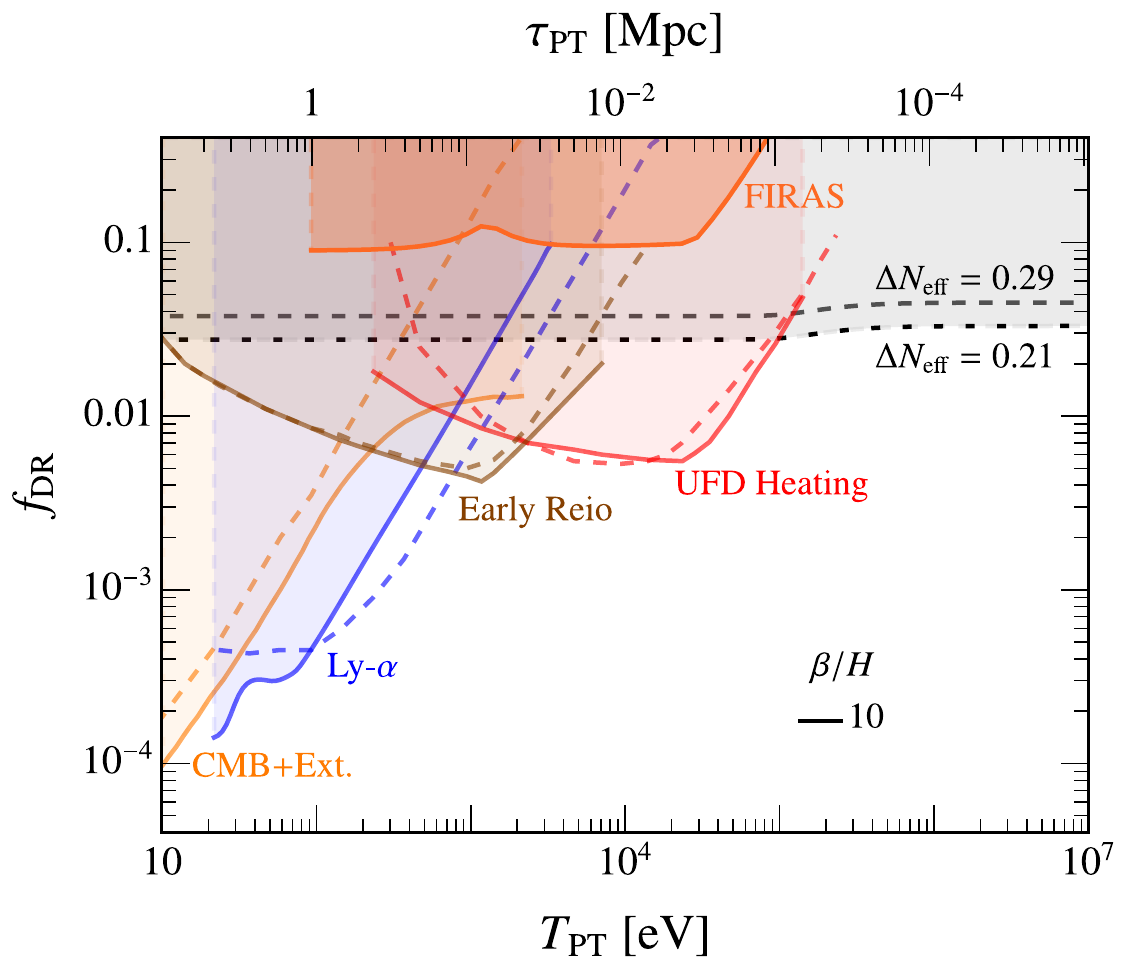}\\\vspace{1mm}\includegraphics[width=0.695\linewidth]{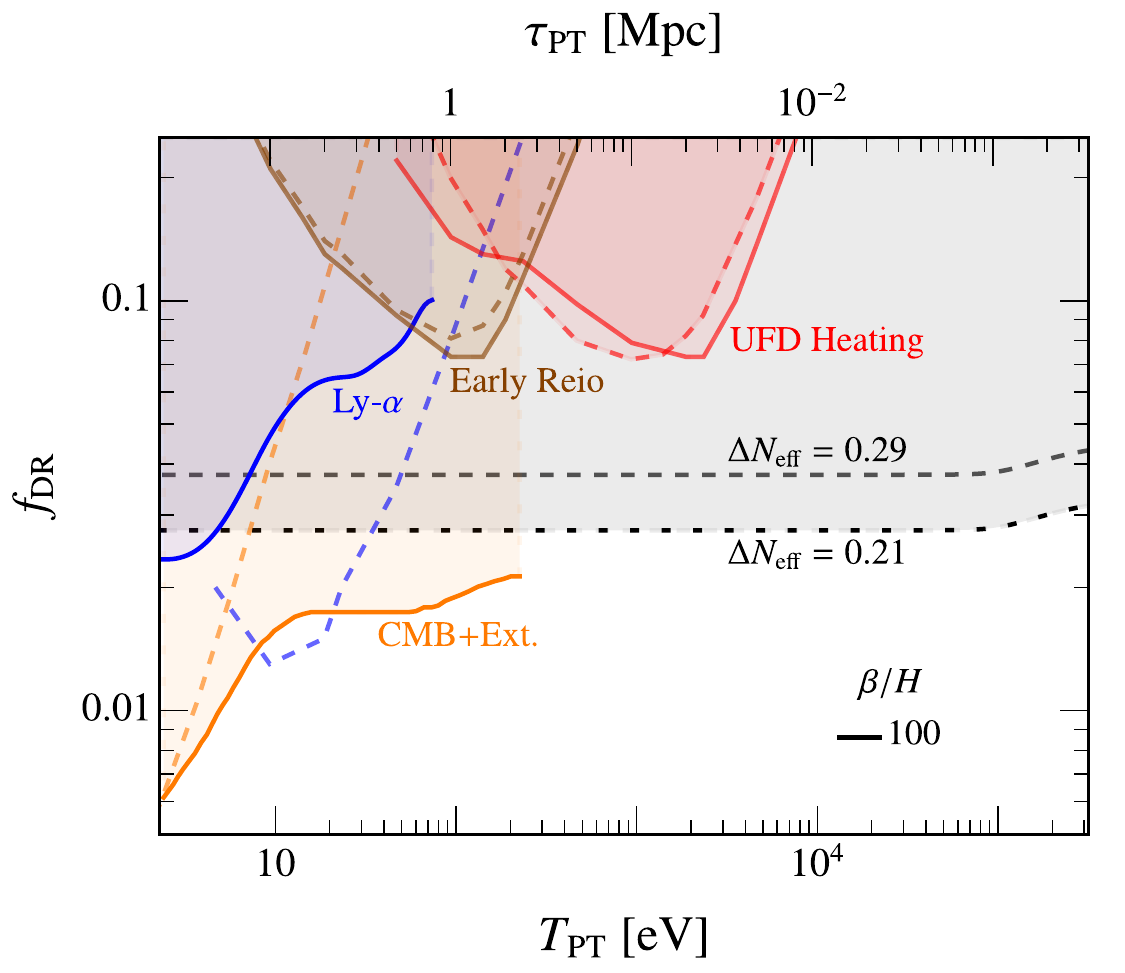}
    \caption{$2\sigma$ exclusion bounds on $f_{\rm DR}$ presented for $\beta/H_{\rm PT}= 10$ (top) and $\beta/H_{\rm PT}= 100$ (bottom). The CMB+Ext. bound (Ext.=BAO+KV450) was calculated using MCMC with the datasets in Refs.~\cite{Planck:2019nip,Planck:2018lbu,Beutler_2011, Ross:2014qpa, BOSS:2016wmc,Hildebrandt:2018yau} (pale orange), where KV450 primarily serves to exclude higher temperature PTs that affect $k$-modes $\sim (0.1-0.3)h/{\rm Mpc}$. The Lyman-$\alpha$ bound was estimated using a reduced $2$-parameter likelihood study~\cite{Bird:2023,Fernandez:2024,He:2025,Bird_github:2026} (blue). We also include bounds calculated from early reionization~\cite{Qin:2025ymc} (brown), UFD galaxy heating~\cite{Graham:2024} (red) and FIRAS~\cite{1994ApJ...420..439M,Fixsen:1996nj} (bright orange). The corresponding bounds derived by direct translation are presented as dashed curves of the same color, using Ref.~\cite{Planck_inf:2020} for CMB+BAO and Ref.~\cite{2011MNRAS.413.1717B} for Lyman-$\alpha$. The existing $\Delta N_{\rm eff}$ constraint (light gray region) includes $\Delta N_{\rm eff} \leq 0.29$ (dashed, TTTEEE+lowE+lensing+BAO) and $\Delta N_{\rm eff} \leq 0.21$ (dotted, TTTEEE+lowE+lensing) from Planck2018~\cite{Planck:2018vyg}.      
        }
    \label{fig:combined_bounds}
\end{figure}

Since ${\cal P}_{\zeta, \rm PT}(k)$ scales with the PT parameters $(z_{\rm PT},\ f_{\rm DR},\ \beta/H_{\rm PT})$ in a well-defined way, bounds on the PT parameters can be obtained by constraining ${\cal P}_{\zeta}(k)$. We draw $2\sigma$ exclusion regions in the $(z_{\rm PT},f_{\rm DR})$ plane for fixed choices of $\beta/H_{\rm PT}$. Fig.~\ref{fig:combined_bounds} summarizes our results for $\beta/H_{\rm PT} = 10$ and $100$. 
For FOPT into DR, a standard constraint comes from a contribution to $\Delta N_{\rm eff}$~\cite{Bai:2022}, that parametrizes any additional radiation component on top of the SM radiation. 
In particular, a constraint on the maximum $\Delta N_{\rm eff}$ places an upper bound on $f_{\rm DR}$ according to
\begin{align}
    f_{\rm DR} 
    = \frac{\Delta N_{\rm eff}}{N_\nu + \Delta N_{\rm eff} + (8/7)(T_\gamma/T_\nu)^4} \,,
\end{align}
where $N_\nu \approx 3.044$~\cite{Froustey:2020mcq,Bennett:2020zkv} is the contribution from the SM neutrinos~\cite{Gariazzo:2019,Bennett:2021,Froustey:2020,Akita:2020}, and $T_\nu/T_\gamma\approx0.716$ is the temperature ratio between the SM neutrinos and photons after the $e^+e^-$ annihilation. 

In Fig.~\ref{fig:combined_bounds}, we summarize the $2\sigma$ bounds on $f_{\rm DR}$ derived in this work, shown for two choices of $\beta/H_{\rm PT}$ and assuming $v_w=1$. The generic constraints on the DR energy density from the Planck~2018 report~\cite{Planck:2018vyg} are shown as dashed lines for $\Delta N_{\rm eff} \leq 0.29$ (TTTEEE+lowE+lensing
+BAO) and dotted lines for $\Delta N_{\rm eff} \leq 0.21$  (TTTEEE+lowE+lensing). Compared to these DR bounds that neglect PT-induced perturbations, most of our constraints derived from the power spectrum ${\cal P}_{\zeta,{\rm PT}}(k)$ are significantly stronger, especially for slower transitions with $\beta/H_{\rm PT}=10$. By combining a range of cosmological observables that probe the curvature spectrum across different length scales, we obtain more stringent bounds over a broad range of horizon scales $\tau_{\rm PT}$, and equivalently, redshifts $z_{\rm PT}$ at which the PT signal is generated. Our bounds are derived using two different approaches:
\\ 
\\
{\bf Translating existing bounds:} We compare the curvature power spectrum ${\cal P}_{\zeta}(k)$ in Eq.~\eqref{eq.totalP} with existing bounds on ${\cal P}_{\zeta}(k)$ from the literature, as cited in the figure caption.
For a given $(z_{\rm PT}, \beta/H_{\rm PT})$, this $\mathcal{P}_\zeta(k)$ comparison yields an estimate of the corresponding constraint on $f_{\rm DR}$. 
To avoid relying on uncertain predictions on length scales smaller than the average bubble size $d_b\approx (8\pi)^{1/3}v_w/\beta$, we truncate the PT spectrum ${\cal P}_{\zeta,\rm PT}(k)$ and retain only modes with $k < a_{\rm PT}/d_b$. 
Bounds obtained using this procedure are shown as dashed curves in Fig.~\ref{fig:combined_bounds}.
However, it is useful to note that the bounds in the literature are derived under different assumptions regarding the shape of ${\cal P}_{\zeta}(k)$: the CMB+BAO~\cite{Planck_inf:2020} and Lyman-$\alpha$~\cite{2011MNRAS.413.1717B} constraints assume a ${\cal P}_{\zeta}(k)$ parameterized by connected knots at discretized $k$-modes, while the early reionization~\cite{Qin:2025ymc} and UFD galaxy bounds~\cite{Graham:2024} assume a delta-function-like bump in $k$. 
\\
\\
{\bf Direct calculation:} We next use the ${\cal P}_{\zeta}(k)$ in Eq.~\eqref{eq.totalP} to compute CMB and structure formation observables directly. The resulting predictions, such as the CMB anisotropy spectra and the halo mass function, are then compared with observational data or existing bounds to derive constraints on $f_{\rm DR}$. A key feature of this analysis is that it explicitly incorporates the robust $k^{3}$ scaling of ${\cal P}_{\zeta,{\rm PT}}(k)$ on large scales, rather than relying on the simplified spectral shapes commonly assumed in the literature when deriving curvature perturbation constraints. Bounds obtained using this procedure are shown as solid curves in both the plots of Fig.~\ref{fig:combined_bounds}.

When directly computing the CMB spectral distortions (Section~\ref{sec:cosmo_distort}), the early reionization (Section~\ref{sec:cosmo_reio}), and the UFD galaxy (Section~\ref{sec:cosmo_UFD}) bounds, we also truncate the PT power spectrum at $k < a_{\rm PT}/d_b$. In contrast, when deriving the CMB anisotropy (Section~\ref{sec:cosmo_CMB}) and Lyman-$\alpha$ (Section~\ref{sec:cosmo_lyman}) bounds using the Boltzmann solver \CLASS~\cite{lesgourgues:2011class,Diego:2011class}, we retain the full peak of ${\cal P}_{\zeta,\rm PT}(k)$ without truncation in order to avoid introducing numerical discontinuities during the parameter scan. Instead, when presenting the CMB anisotropy and Lyman-$\alpha$ results, we impose a lower cutoff on $T_{\rm PT}$ since for even lower temperatures, the data will be probing the scales smaller or comparable to the bubble scale, which is where our ${\cal P}_{\zeta, {\rm PT}}(k)$ computation is less accurate. 
For $\beta/H_{\rm PT}= 10$, we therefore consider $T_{\rm PT}\geq 10$~eV for the CMB bound and $T_{\rm PT}\geq 21$~eV for the Lyman-$\alpha$ bound. 
For $\beta/H_{\rm PT} = 100$, the bubble size is ten times smaller at fixed $T_{\rm PT}$, such that the sub-bubble regime extends to $k$ scales that are ten times larger with temperature cutoffs ten times lower.

We make a few remarks regarding the constraints shown in Fig.~\ref{fig:combined_bounds}. First, the bounds derived directly from the early reionization (brown) and UFD galaxy observations (red) are comparable to, but can be slightly stronger than, those obtained by translating existing curvature spectrum limits assuming a delta-function feature. This improvement arises from properly accounting for the $k^{3}$ infrared scaling of the PT–induced curvature spectrum, which enhances the halo mass function relevant for these constraints.

Second, consistent with Ref.~\cite{Elor:2023xbz}, the CMB spectral distortion bound from FIRAS is weaker than the $\Delta N_{\rm eff}$ constraint. For the curvature perturbations considered here, these spectral distortion measurements are therefore not currently the leading probe. However, away from the supercooled phase-transition regime (i.e., for larger $\beta/H_{\rm PT}$), small-scale inhomogeneities sourced by sound waves, which are neglected in this analysis, may generate large density perturbations and significantly strengthen the spectral distortion constraints~\cite{Ramberg:2022irf}.

The CMB anisotropy constraints (orange) are generally stronger than the homogeneous $\Delta N_{\rm eff}$ bounds obtained without including PT-induced perturbations. Because the ${\cal P}_{\zeta,{\rm PT}}(k)$ falls off as $k^3$ on large scales, earlier PTs generate a smaller perturbation signal within the CMB sensitivity range probed by {\it Planck}, $k \lesssim 0.1~{\rm Mpc}^{-1}$. We therefore expect the CMB bounds from our MCMC analysis (solid orange) to asymptotically approach the $\Delta N_{\rm eff}$ constraint. In the upper panel of Fig.~\ref{fig:combined_bounds}, we observe a modest deviation from the $\Delta N_{\rm eff}$ bound for $\beta/H_{\rm PT}=10$ at $T_{\rm PT}\simeq10^{3}$ eV. In contrast, for $\beta/H_{\rm PT}=100$ (lower panel), the constraints converge at high $z_{\rm PT}$ to the {\it Planck} limit better. This behavior reflects the diminishing impact of PT-induced perturbations on the CMB once ${\cal P}_{\zeta,{\rm PT}}(k)$ peaks at higher $k$-modes with larger $\beta/H_{\rm PT}$. Notice that since we also include KV450 data in the MCMC analysis, which plays a more important role in constraining PT with $T_{\rm PT}\gtrsim 10^2$~eV, the CMB+Ext. bound we show in the plot still remains stronger than the Planck $\Delta N_{\rm eff}$ bounds.

Finally, the Lyman-$\alpha$ forest (blue) bounds derived by translating existing constraints (dashed) should be regarded as approximate since those analyses, where the bounds were taken from, assume primordial shapes that are parameterized differently (with a different number of free parameters) than the PT-induced spectrum ${\cal P}_{\zeta}(k)$ that we apply the bounds to. The bounds obtained from fits to the compressed data (solid) should similarly be regarded as an estimate due to the limitations of the Lyman-$\alpha$ calculation that will be discussed in Section~\ref{sec:cosmo_lyman}. 

Overall, PT-induced curvature perturbations enable strong constraints on $f_{\rm DR}$ for long-duration transitions like $\beta/H_{\rm PT}=10$, but the constraining power drops steeply as a function of $\beta/H_{\rm PT}$.
Equation \eqref{eq:Pdt} demonstrates that $\mathcal{P}_{\delta t} \propto (\beta/H_{\rm PT})^{-5}$, such that going from $\beta/H_{\rm PT}$ = 10 to 100 suppresses the amplitude by a factor of $10^5$, which would require $f_{\rm DR}$ to increase by $10^{2.5}$ to compensate. However, even for the $\beta/H_{\rm PT}=100$ case (bottom panel of Fig.~\ref{fig:combined_bounds}), measurements from CMB anisotropy still provide a more stringent constraint compared to $\Delta N_{\rm eff}$.
In the following, we provide further details on the derivation of the above bounds.

\section{Constraints from the cosmic microwave background}\label{sec:CMB}
Enhancements to the curvature spectrum ${\cal P}_{\zeta}(k)$ will enhance the photon density and metric perturbations before recombination, leading to measurable modifications to the CMB temperature and polarization spectra, and producing $\mu$- and $y$-type distortions of the CMB.

\subsection{MCMC analysis using CMB and BAO data}
\label{sec:cosmo_CMB}
CMB measurements set an upper bound on the primordial curvature power spectrum~\cite{Planck_inf:2020}. By applying this bound directly to the PT enhanced spectrum ${\cal P}_{\zeta}(k)$, we can translate the bound into approximate constraints on the PT parameters~\cite{Elor:2023xbz}. These bounds are presented as the dashed pale orange curves in Fig.~\ref{fig:combined_bounds}.  

For a more careful analysis of the effect that CMB constraints have on the PT parameters, we conduct a likelihood study using an MCMC analysis with CMB and BAO datasets. We use \CLASS\ to solve the full Einstein-Boltzmann equations for the matter and radiation perturbations. We implement the primordial spectrum ${\cal P}_{\zeta}(k)$ from \eqref{eq.totalP} through the built-in \texttt{external\_pk} module. A total of six parameters were used in \texttt{external\_pk}: three $\Lambda$CDM parameters $(A_s,\ n_s,\ k_{\rm pivot})$ for ${\cal P}_{\zeta, ad}(k)$ and three PT parameters $(z_{\rm PT},\ f_{\rm DR},\ \beta/H_{\rm PT})$ for ${\cal P}_{\zeta, \rm PT}(k)$. As mentioned earlier, the full ${\cal P}_{\zeta, \rm PT}(k)$ peak without truncations (at the bubble size scale) was included to avoid introducing any discontinuities to the numerical calculations during the parameter scan. The DR abundance from the PT is also accounted for in \CLASS\ by adding it to the budget of massless free-streaming neutrinos:   
\begin{equation}
    \tt \Omega_{ur} \longrightarrow  \Omega_{ur} + \frac{f_{\rm DR}}{1-f_{\rm DR}}\left(\Omega_{ur} + \Omega_{g}\right) = \frac{\Omega_{ur} + f_{\rm DR}*\Omega_g}{1-f_{\rm DR}}
\end{equation}
where $\tt \Omega_{ur}$ and $\tt \Omega_{g}$ are the \CLASS\ density parameters for the massless neutrinos and photons respectively\footnote{In this work, we include only the modification of the curvature perturbation $\mathcal{P}_\zeta(k)$ induced by the PT and do not consider a separate DR isocurvature mode. The PT contribution to $\mathcal{P}_\zeta(k)$ affects the initial perturbations of all cosmic fluids and the metric, and is suppressed by $f_{\rm DR}^{2}$ \eqref{eq.totalP}. Likewise, the initial photon and neutrino perturbations sourced by a DR isocurvature mode are also suppressed by $f_{\rm DR}^{2}$~\cite{Ghosh:2021axu,Chang:2025uvx}. We therefore expect the curvature and isocurvature contributions from the PT to produce cosmological signals of comparable magnitude and would not change the bounds significantly. A detailed analysis of the isocurvature signal is left for future work.}.

A subtlety in implementing the modified primordial spectrum is that perturbation modes entering the horizon before the PT should, in principle, receive growth factor corrections. However, since photon density perturbations undergo diffusion damping inside the horizon, delays in the sub-horizon evolution for a given perturbation $k$-mode would imply less damping compared to $\Lambda$CDM. 
By treating $\mathcal{P}_{\zeta,{\rm PT}}(k)$ as a primordial initial condition, modes that would not acquire PT-sourced perturbations until the PT actually takes place at $\tau_{\rm PT}$ are instead subjected to Silk damping effects from the time they first enter the horizon. This overestimates the damping of the PT signal, making our CMB bounds conservative.

\begin{figure}[t!]
    \centering
    \includegraphics[width=0.5\linewidth]{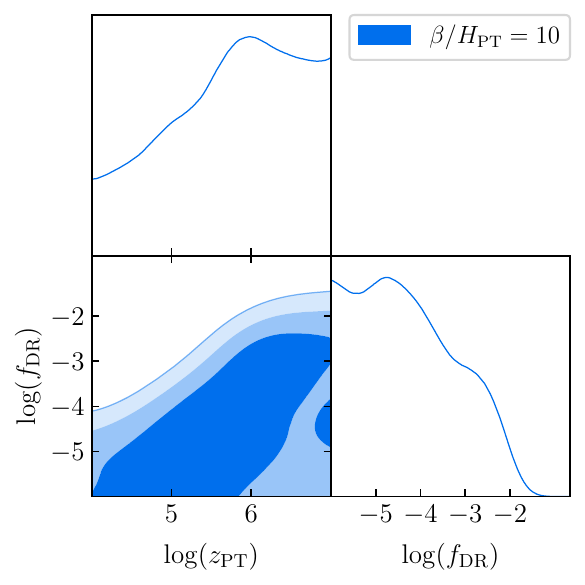}\includegraphics[width=0.5\linewidth]{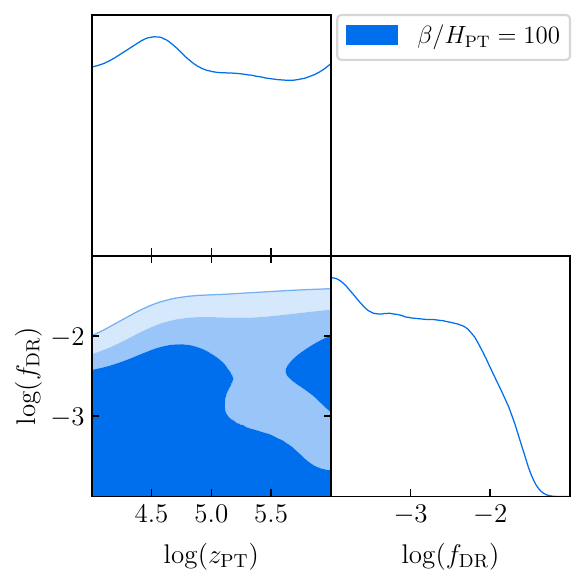}
    \caption{Results of MCMC analysis with CMB + BAO +KV450 data for $\beta/H_{\rm PT}= 10$ (left) and $\beta/H_{\rm PT}= 100$ (right), marginalized to $z_{\rm PT}$ and $f_{\rm DR}$ in log space (base $10$). 2D contour plots show the  preferred regions up to 1, 2 and 3$\sigma$ boundaries. The MCMC scan was done with $\Delta N_{\rm eff}$ component varying with $f_{\rm DR}$, which contributes to the suppressed preference for large $f_{\rm DR}$ values.}
    \label{fig:CMB-triangle}
\end{figure}

The parameter space of the scan consists of the usual six $\Lambda$CDM parameters and two additional PT parameters: $z_{\rm PT}$ and $f_{\rm DR}$. The cosmological parameters and their prior ranges are listed in Table \ref{tab:prior}. 
\begin{table}
    \centering
    \begin{tabular}{|c|c|}\hline
       $\Lambda$CDM Parameters  & Range  \\\hline
       $10^2 \omega_{\rm b}$    &$[1.8, 3]$      \\\hline
       $\omega_{\rm cdm}$       &$[0.1, 0.2]$    \\\hline
       $H_0 \rm[km/s/Mpc]$      &$[60.0, 80.0]$  \\\hline
       $\tau_{\rm reio}$        &$[0.004, 0.12]$ \\\hline
       $10^9A_s$                &$[1.8, 3]$      \\\hline
       $n_s$                    &$[0.9, 1.1]$    \\\hline
    \end{tabular}
    
    \vspace{0.3 cm}
    
    \begin{tabular}{|c|c|c|}\hline
       PT Parameters  & Range ($\beta/H_{\rm PT}= 10$)  & Range ($\beta/H_{\rm PT}= 100$) \\\hline
       $\log_{10}(z_{\rm PT})$  &$[4, 7]$        & $[4, 6]$\\\hline
       $\log_{10}(f_{\rm DR})$  &$[-6, 0]$       & $[-4, -1]$\\\hline
    \end{tabular}
    \caption{Cosmological parameters and their prior ranges for MCMC analysis. For the PT parameters, we fix $\beta/H_{\rm PT}$  and scan over $z_{\rm PT}$ and $f_ {\rm DR}$ in log space (base 10). The same $\Lambda$CDM priors were used in both cases.}
    \label{tab:prior}
\end{table}
We use a combination of the following datasets for our MCMC analysis.
 \begin{itemize}
     \item \underline{Planck}: Measurements of CMB temperature and polarization spectra from the {\it Planck} 2018 dataset, which consists of the low-$\ell$ $(\ell < 30)$ TT, EE and the high-$\ell$ $(\ell \geq 30)$ TTTEEE measurements~\cite{Planck:2019nip}. It also includes the {\it Planck} Lensing likelihood~\cite{Planck:2018lbu}.
     \item \underline{BAO}: Baryon Acoustic Oscillation (BAO) measurements, which consists of the 6DF Galaxy survey, SDSS-DR7 MGS data, and the BOSS measurement of the BAO scale and $f\sigma 8$ from the DR12 galaxy sample~\cite{Beutler_2011, Ross:2014qpa, BOSS:2016wmc}.
     \item \underline{KV450}: The KV450 dataset consists of the matter power spectrum shape measurement from KiDS + Viking 450~\cite{Hildebrandt:2018yau}. In the KV450 likelihood, we use data up to $k_{\rm max} = 0.3h ~{\rm Mpc}^{-1}$ to remain in the linear regime. We cross check the bounds with only CMB+BAO data, and the inclusion of KV450 only modifies the bound significantly when $T_{\rm PT}\gtrsim 10^2$~eV. For PTs at these higher temperature, the $k_{\rm max}$ we use only probes the $\mathcal{P}_{\zeta,{\rm PT}}(k)$ spectrum on superhorizon scales $k < 1/\tau_{\rm PT}$, and it is safe not to include the subhorizon growth function correction. The inclusion of the KV450 also strengthens the bound at higher $T_{\rm PT}$ compared to the Planck $\Delta N_{\rm eff}$ constraints.
 \end{itemize}
The MCMC analysis was performed using \texttt{MontePython}~\cite{Audren:2012wb,Brinckmann:2018cvx} with the Metropolis–Hastings algorithm~\cite{Hastings:1970aa}. We adopt the convention in~\cite{Planck:2018vyg} of modeling the free-streaming neutrinos as two massless species and one massive species with mass $0.06$ eV, with $f_{\rm DR}$ modifications to the massless component as described in Section \ref{sec.analysis}. The Gelman-Rubin convergence criteria $R-1 < 0.05$ was satisfied for the scans \cite{Rubin:1992} and have verified that the 2$\sigma$ contours are stable under further sampling. We use \texttt{GetDist} to analyze and plot the MCMC samples~\cite{Lewis:2019xzd}. In Fig.~\ref{fig:CMB-triangle}, we show the preferred regions for $\beta/H_{\rm PT}= 10$ and $\beta/H_{\rm PT}= 100$ marginalized to $z_{\rm PT}$ and $f_{\rm DR}$, with the 2D contours in $z_{\rm PT}$ and $f_{\rm DR}$ plotted up to $3 \sigma$. 

The $2\sigma$ boundaries from Fig.~\ref{fig:CMB-triangle} were extracted and presented as the solid pale orange exclusion bounds in Fig.~\ref{fig:combined_bounds}.

\subsection{CMB spectral distortion}
\label{sec:cosmo_distort}
In the early Universe, energy stored in curvature perturbations with $k \lesssim 5400\,\mathrm{Mpc}^{-1} $ is dissipated by Silk damping, producing 
$\mu$- and $y$-type distortions of the CMB. For modes with 
$ k_{\rm min} \gtrsim 1\,\mathrm{Mpc}^{-1} $, the photon-baryon fluid is tightly 
coupled and baryon loading can be neglected. In this regime, the resulting 
spectral distortions can be approximated as~\cite{Chluba:2012we}
\begin{eqnarray}
|\mu|~{\rm or}~|y| \simeq A \int_{k_{\rm min}}^\infty \frac{\mathrm{d}k}{k}\,
\mathcal{P}_\zeta(k)
\left[
B\, e^{-k/(5400\,\mathrm{Mpc}^{-1})}
- C\, e^{-(k/(31.6\,\mathrm{Mpc}^{-1}))^2}
\right],
\end{eqnarray}
where $ k_{\rm min} = 1\,\mathrm{Mpc}^{-1} $, with 
$(A,B,C)_\mu=(2.2,1,1)$ and $(A,B,C)_y=(0.4,0,-1)$. 

Comparing these expressions with the FIRAS limits 
$ |\mu| < 9.0\times10^{-5} $ and $ |y| < 1.5\times10^{-5}$~\cite{1994ApJ...420..439M,Fixsen:1996nj}, we obtain the exclusion region as the bright orange curve in Fig.~\ref{fig:combined_bounds}. As the PT  temperature $T_{\rm PT}$ decreases, the $y$-distortion constraint becomes more stringent than the $\mu$-distortion bound at $T_{\rm PT} \simeq 10^3~(10^2)\,\mathrm{eV}$ for $\beta/H_{\rm PT}=10~(100)$. We set the lower integration limit to $k=1,{\rm Mpc}^{-1}$, below which the tightly coupled photon–baryon approximation starts to break down and a more careful treatment of photon–baryon perturbations is required.

\section{Structure formation constraints}\label{sec:Struct}
Enhancements to the curvature spectrum ${\cal P}_{\zeta}(k)$ also enhance the initial DM perturbations on the relevant scales, which in turn impact galaxy cluster and halo formation. Measurements of structure formation can thus be used to set constraints on ${\cal P}_{\zeta}(k)$. 

A subtlety in estimating the structure formation bound is that matter perturbations already grow logarithmically after entering the horizon during the radiation era. 
As discussed in Sec.~\ref{sec.analysis}, for $k$-modes that are already within the horizon when the PT occurs, it is important to apply a growth factor correction.
Specifically, consider a matter perturbation mode $k$ that enters the horizon at $1/k<\tau_{\rm PT}$ before the PT. In $\Lambda$CDM cosmology, the power in the mode at conformal time $\tau$ after horizon entry is $\Delta_{\rm DM}(1/k,\tau)^2\times{\cal P}_{\zeta}(k)$, where $\Delta_{\rm DM}(\tau_i,\tau)$ is the matter growth factor from $\tau_i$ to $\tau$. The ${\cal P}_{\zeta, \rm PT}(k)$-induced component of the mode, however, is not generated until $\tau_{\rm PT}$, so the initial $\tau_i$ of the sub-horizon evolution has to be delayed till $\tau_{\rm PT}$ to obtain a reduced power $\Delta_{\rm DM}(\tau_{\rm PT},\tau)^2\times{\cal P}_{\zeta, \rm PT}(k)$ for this component. 
Therefore, when calculating an observable at comoving time $\tau$, we take into account the following growth factor correction using a rescaled ${\cal P}_{\zeta, \rm PT}(k)$:
\begin{equation}\label{eq:Pzeta_trans}
    {\cal P}_{\zeta, \rm PT}(k) \longrightarrow {\cal P}_{\zeta, \rm PT}(k)\times\left(\frac{\Delta_{\rm DM}(\tau_{\rm PT},\tau)}{\Delta_{\rm DM}(1/k,\tau)}\right)^2, \qquad\quad ({\rm for}\,\,k>1/\tau_{\rm PT})
\end{equation}
before adding it to the total ${\cal P}_{\zeta}(k)$ when studying structure formation constraints. 

\subsection{Lyman-$\alpha$ bound}
\label{sec:cosmo_lyman}
Measurements of the Lyman-$\alpha$ forest provide constraints on the primordial curvature perturbations. Given the PT enhanced curvature spectrum ${\cal P}_{\zeta}(k)$ with the growth factor correction in \eqref{eq:Pzeta_trans}, the existing Lyman-$\alpha$ bound in Ref.~\cite{2011MNRAS.413.1717B} can be directly applied to estimate constraints on the PT parameters~\cite{Elor:2023xbz}. 
The Lyman-$\alpha$ bounds obtained in this way are presented as the dashed blue curves in Fig.~\ref{fig:combined_bounds}.  

A more careful analysis of the bounds would require a calculation of the effect of our particular ${\cal P}_{\zeta}(k)$ spectrum on the Lyman-$\alpha$ forest flux power spectrum. Instead of performing the full suite of Lyman-$\alpha$ calculations, which would involve computationally heavy hydrodynamical simulations, we estimate the Lyman-$\alpha$ constraints on the PT parameters by conducting a simplified likelihood study using a compressed Lyman-$\alpha$ likelihood. Our approach is to compute the linear matter power spectrum at each point in the PT parameter space using \texttt{CLASS}, apply the growth factor correction at the Lyman-$\alpha$ pivot scale, and then compare against the compressed eBOSS data at that scale. The data consists of two parameters~\cite{McDonald:2000,Chabanier:2019,Pedersen:2020,Pedersen:2021,Pedersen:2023,Goldstein:2023}:   
\begin{equation}
\Delta^2_{\rm lin} \equiv \frac{k^3_p}{2\pi^2} {\rm P}_{\rm lin}(k_p,z_p), \qquad\qquad n_{\rm lin} \equiv \left.\frac{{\rm d}\ln {\rm P}_{\rm lin}(k,z)}{{\rm d}\ln k}\right|_{k_p,z_p},
\end{equation}
where $\Delta^2_{\rm lin}$ is the amplitude and $n_{\rm lin}$ the tilt of the linear matter power spectrum ${\rm P}_{\rm lin}(k,z)$, evaluated at a pivot redshift $z_p = 3$ and pivot scale $0.009\ {\rm s/km}$ in velocity units, which translates to the conformal scale $k_p \approx 1.03{h\rm~   Mpc}^{-1}$~\cite{Bagherian:2024,Bansal:2024,Buckley:2025}. The dataset we use for this reduced two-parameter study is: 
\begin{itemize}
    \item \underline{PRIYA}: a compressed likelihood using the 1D Lyman-$\alpha$ flux power spectrum measurements from SDSS DR14 BOSS and eBOSS quasars~\cite{Bird:2023,Fernandez:2024,He:2025}. Following Ref.~\cite{Bird_github:2026}, we use $\Delta^2_{\rm lin} = 0.267 \pm 0.022$ and $n_{\rm lin} = -2.288 \pm 0.024$ with correlation coefficient $0.4$ to construct a 2D Gaussian likelihood.
 \end{itemize}
Since the compressed likelihood requires the derivative of the matter power spectrum ${\rm P}_{\rm lin}(k,z)$ to be well-defined at $k_p$, we require the primordial power spectrum ${\cal P}_{\zeta}(k)$ to be smooth with no sharp discontinuities. As discussed in Section~\ref{sec.analysis}, we therefore include the full peak of ${\cal P}_{\zeta, \rm PT}(k)$ for this study.

Limitations in accounting for growth factor corrections prevent us from running a full, accurate MCMC analysis of ${\cal P}_{\zeta}(k)$ with this compressed Lyman-$\alpha$ likelihood. Using the same \CLASS\ implementation of ${\cal P}_{\zeta}(k)$ and DR abundance as was used in Section~\ref{sec:cosmo_CMB}, we instead adopt a direct approach of calculating ${\rm P}_{\rm lin}(k,z)$ and the \CLASS\ CDM \textit{transfer function} $T_{\rm DM}(k,\tau)$ to perform the $\chi^2$-fit. In doing the \CLASS\ calculations, the $\Lambda$CDM parameters were held fixed while the PT parameters $z_{\rm PT}$ and $f_{\rm DR}$ were varied for fixed $\beta/H_{\rm PT}$ \footnote{The results obtained are thus only an estimate of the Lyman-$\alpha$ constraint since the $\Lambda$CDM parameters would also be allowed to vary in a full MCMC analysis.}. The values and ranges used are shown in Table \ref{tab:class}.
\begin{table}
    \centering
    \begin{tabular}{|c|c|}\hline
       \CLASS\ $\Lambda$CDM Parameters  & Fixed                   \\\hline
       $\omega_{\rm b}$                 &$0.02238280$             \\\hline
       $\omega_{\rm cdm}$               &$0.1201075$              \\\hline
       $h$                              &$0.67810$                \\\hline
       $\tau_{\rm reio}$                &$0.05430842$             \\\hline
       $A_s$                            &$2.100549\times 10^{-9}$ \\\hline
       $n_s$                            &$0.9660499$              \\\hline
       $k_{\rm pivot}~[{\rm Mpc}^{-1}]$ &$0.05$                   \\\hline
    \end{tabular}
    
    \vspace{0.3 cm}
    
    \begin{tabular}{|c|c|c|}\hline
       PT Parameters  & Range ($\beta/H_{\rm PT}= 10$)  & Range ($\beta/H_{\rm PT}= 100$) \vspace{0.3mm}\\\hline
       $z_{\rm PT}$  &$[2\times 10^4, 4\times 10^7]$& $[2\times 10^3, 7\times 10^5]$ \vspace{0.3mm}\\\hline
       $f_{\rm DR}$  &$[10^{-4}, 10^{-1}]$       & $[10^{-2}, 10^{-1}]$ \\\hline
    \end{tabular}
    \caption{\CLASS\ parameter settings and ranges for compressed Lyman-$\alpha$ 2 parameter study of linear matter power spectrum. The $\Lambda$CDM parameters were held fixed while the PT parameters were looped over in log spaced steps (base 10). Note that $k_{\rm pivot}$ denotes a \CLASS\ parameter for the $\Lambda$CDM ${\cal P}_{\zeta, \rm ad}(k)$ spectrum, which should not be confused with $k_p$ for the Lyman-$\alpha$ scale where the compressed data applies.}
    \label{tab:class}
\end{table}
Taking $k_p$ as the mode that we observe in this likelihood study, we apply the \CLASS\ transfer function to ${\rm P}_{\rm lin}(k,z)$ such that if $k_p$ entered the horizon before $\tau_{\rm PT}$, then:   
\begin{eqnarray}\label{eq:Plin_trans}
    {\rm P}_{\rm lin}(k,z_p) = {\rm P}_{\rm lin, ad}(k,z_p) + \left(\frac{T_{\rm DM}(k,1/k_p)}{T_{\rm DM}(k,\tau_{\rm PT})}\right)^2 {\rm P}_{\rm lin, PT}(k,z_p) \quad\quad (k_p > \tau_{\rm PT}^{-1})
\end{eqnarray}
% \yt{relate this to (5.1)} 
where ${\rm P}_{\rm lin, ad}(k,z_p)$ is the $\Lambda$CDM matter power spectrum following from ${\cal P}_{\zeta, \rm ad}(k)$, while ${\rm P}_{\rm lin, PT}(k,z_p)$ denotes the PT induced component sourced from ${\cal P}_{\zeta, \rm PT}(k)$. For reference, the \CLASS\ transfer function $T_{\rm DM}(k,\tau)$ is calculated for a mode $k$ evaluated at conformal time $\tau$, and is related to the growth factor in Eq.~\eqref{eq:Pzeta_trans} by
\begin{equation}
    \Delta_{\rm DM}(\tau_i,\tau) = \frac{T_{\rm DM}(k,\tau)}{T_{\rm DM}(k,\tau_i)}.
\end{equation}
The transfer function correction in Eq.~\eqref{eq:Plin_trans} is thus equivalent to applying $1/\Delta_{\rm DM}(1/k_p,\tau_{\rm PT})^2$ to delay the sub-horizon growth of PT-induced matter perturbations at the scale $k_p$ to $\tau_{\rm PT}$\footnote{We nonetheless need a small range of $k$-modes in the vicinity of $k_p$ for taking the derivative, for which we calculate several $T_{\rm DM}(k,\tau)$ for $k$-modes in the range $k=k_p\pm dk$ where $dk = 0.0005/$Mpc. At a given $\tau$, we then  interpolate in $k$ across this range.}. For each choice of parameters, the goodness of fit was then determined by taking $\Delta\chi^2$ with respect to $\Lambda$CDM as the best fit model:
\begin{equation}
    \Delta\chi^2 = \chi_{\rm PT+ad}^2 - \chi_{\rm ad}^2
\end{equation}
where $\chi^2_{\rm ad}$ is the $\chi^2$-fit for ${\rm P}_{\rm lin, ad}(k,z_p)$ only and $\chi_{\rm PT+ad}^2$ is the $\chi^2$-fit of the full ${\rm P}_{\rm lin}(k,z_p)$ including growth factor corrections. Since the fit was done by varying $f_{\rm DR}$ for each $z_{\rm PT}$, we keep points within $\Delta\chi^2 \leq 3.84$ ($2\sigma$ for 1 degree of freedom).

The $2\sigma$ boundary curves are presented as the solid blue curves in Fig.~\ref{fig:combined_bounds}. Compared to the dashed blue curves, the calculated Lyman-$\alpha$ bound is noticeably weaker for $\beta/H_{\rm PT} =100$, and in the high $z_{\rm PT}$ region for $\beta/H_{\rm PT}= 10$. Since we were performing a reduced Lyman-$\alpha$ likelihood study, this feature may be due to the limitations of constraining only a single perturbation mode $k_p$ in the matter power spectrum. As the peak of the PT signal moves away from $k_p$ to higher $k$-modes, whether by increasing $z_{\rm PT}$ or by increasing $\beta/H_{\rm PT}$  for the same $z_{\rm PT}$, there would be an expected fall-off in constraining power. As we observe in the plot for $\beta/H_{\rm PT}= 10$, the solid curve gives a weaker bound than the dashed curve until $z_{\rm PT}$ decreases to the scale $\tau_{\rm PT} \sim 1\ {\rm Mpc} \sim 1/k_p$, after which it gives a stronger constraint. A visible dip in the bounds for $\beta/H_{\rm PT}= 10$ is also apparent towards low $z_{\rm PT}$. This feature occurs when $k_p$ approaches the peak of ${\cal P}_{\zeta, \rm PT}(k)$ and the bound curves upwards again as it passes the peak. We nonetheless cutoff the bound below $T_{\rm PT} \approx 21$~eV where $k_p$ probes ${\cal P}_{\zeta, \rm PT}(k)$ on length scales smaller than the bubble scale.  

\subsection{Early reionization}
\label{sec:cosmo_reio}
We follow the calculation in~\cite{Kumar:2025} to derive the early reionization constraint. The basic idea is that if a FOPT produces too large matter perturbations in the early universe, structure will form too early. The ionizing radiation from the first stars and galaxies would then reionize the universe earlier than indicated by the CMB~\cite{Planck:2018vyg} and by observations of absorption features in the spectra of distant quasars~\cite{Becker:2014oga}. This effect allows us to constrain the size of the primordial perturbations by requiring that structure formation at high redshift does not become too efficient. 

Using this approach, Ref.~\cite{Kumar:2025} derives an upper bound on the primordial curvature perturbation assuming a narrow peak at a given $k$-mode. By comparing their bound with the curvature perturbation generated by the phase transition, after including the transfer-function correction discussed at the start of Section~\ref{sec:Struct}, we obtain the PT constraint shown as the brown dashed curve in Fig.~\ref{fig:combined_bounds}.

However, the curvature perturbation produced by a FOPT is not delta-function–like, but instead exhibits a $k^{3}$ spectrum on large scales. It is therefore important to assess whether the excess power relative to a narrow peak modifies the resulting constraint. To address this, we recast the bound by repeating the analysis of Ref.~\cite{Kumar:2025} using the PT power spectrum $\mathcal{P}_\zeta$. We first perform a standard Press–Schechter calculation to estimate the halo mass function
\begin{equation}\label{eq.dndM}
\frac{{\rm d}n}{{\rm d}M}=\frac{\rho_m}{M}\sqrt{\frac{2}{\pi}}\frac{\delta_c}{\sigma^2}\exp\left(-\frac{\delta_c^2}{2\sigma^2}\right)\frac{{\rm d}\sigma}{{\rm d}M}\,.
\end{equation}
Here $n$ is the comoving number density of collapsed objects as a function of the object mass $M$. $\rho_m\approx 4\times 10^{10}$~$M_{\odot}\,$Mpc$^{-3}$ is today's matter density, $\delta_c\approx 1.68$ is the critical density contrast, and $\sigma^2$ is the variance of density perturbation smoothed over a sphere of size $R(M)= \left(\frac{3M}{4\pi\rho_{\rm m}}\right)^{1/3}$:
\begin{equation}\label{eq.sig2}
\sigma^2(M,z)=\displaystyle{\int}_0^{\infty}\frac{{\rm d}k}{k}\mathcal{P}_\zeta(k)\,T_{\rm DM}^2(z,k) W^2(k\,R(M))\,.
\end{equation}
Using the curvature power spectrum $\mathcal{P}_\zeta(k)$ in Eq.~(\ref{eq.totalP}), we rescale the PT spectrum with the growth factor given in Eq.~(\ref{eq:Pzeta_trans}) up to matter-radiation equality, $\tau=\tau_{\rm eq}\simeq110$~Mpc. Because the perturbation modes relevant for the halo mass function enter the horizon deep in the radiation-dominated era, we remove the logarithmic growth of matter perturbations from horizon entry, $\tau\simeq1/k$, to the PT time $\tau_{\rm PT}$. Treating these modes as purely primordial would otherwise overestimate their growth.

In this regime, the growth factor $\Delta_{\rm DM}(\tau_i,\tau_{\rm eq})$ can be approximated analytically~\cite{dodelson:2020,Hu:1996} as
\begin{equation}\label{eq:trans_approx}
\Delta_{\rm DM}(\tau_i,\tau_{\rm eq}) \simeq 6.4\,\ln\left(0.44\,\tau_{\rm eq}/\tau_i\right),
\end{equation}
where $\tau_i=1/k$ or $\tau_{\rm PT}$. We apply the same growth factor correction when translating the curvature perturbation bound from Ref.~\cite{Kumar:2025}, giving the dashed brown curve in Fig.~\ref{fig:combined_bounds}. After obtaining the modified curvature spectrum, we compute the resulting dark matter perturbations using the transfer function $T_{\rm DM}(z,k)$ from \CLASS\, adopting a top-hat window function $W(x)=3x^{-3}(\sin x - x\cos x)$ throughout the calculation.

A commonly used measure of structure formation in simple reionization and star-formation models is the collapse fraction,
\begin{equation}
f_{\rm coll}(z,M_{\rm min})=\frac{1}{\rho_m}\int_{M_{\rm min}}^\infty {\rm d}M\,M\frac{{\rm d}n}{{\rm d}M}(z)\,,
\end{equation}
which represents the fraction of matter collapsed into halos above a minimum mass $M_{\rm min}$. To be conservative, we focus on very early reionization at $1+z=20$ and include halos with masses above $M_{\rm min}=2\times10^{7}M_{\odot}$. This mass threshold ensures that the halo virial temperature exceeds $10^{4}$~K, allowing efficient atomic cooling while suppressing molecular cooling. As a result, dense dark matter halos form and accrete baryons, with the associated radiation ionizing hydrogen atoms. Following the same assumption for the ionization efficiency as in~\cite{Kumar:2025}, we require
\begin{equation}
f_{\rm coll}(z,2\times 10^7~M_{\odot})<0.1
\end{equation}
to set an upper bound on $f_{\rm DR}$ for given $\beta/H_{\rm PT}$  and $z_{\rm PT}$. The resulting constraint is shown as the solid brown curves in Fig.~\ref{fig:combined_bounds}. The early-reionization bound fills the gap between the Ly-$\alpha$ and UFD heating constraints (discussed below), providing important coverage of intermediate PT scenarios.

\subsection{Ultra-faint dwarf galaxies}
\label{sec:cosmo_UFD}

Following Refs.~\cite{Graham:2023unf,Graham:2024hah}, a large primordial power spectrum can significantly enhance the formation of small-scale dark matter structure, leading to the production of dark matter clumps with masses well above stellar scales. The gas associated with these dark matter objects is typically hotter than stellar gas and can therefore heat the stellar component through purely gravitational interactions and cause the expansion of the stellar scale radius. Measurements of the half-light radii of UFD galaxies, such as Segue-I~\cite{2018ApJ...860...66M}, thus constrain the abundance of such dark matter substructures. These constraints can then be translated into an upper bound on the primordial power spectrum. 

Following this approach, Ref.~\cite{Graham:2024hah} estimates the bound on the primordial power spectrum with the shape of a delta-function peaked at a certain $ k$-mode. In Fig.~\ref{fig:combined_bounds}, we compare the PT power spectrum to their bound, for the case of top-hat window function and $\Delta=200$, to obtain the $f_{\rm DR}$ bound as the red-dashed curve. Here $\Delta$ denotes the assumed overdensity of a DM clump relative to the ambient matter density resulting from gravitational collapse.

Similar to the early reionization constraint, we also want to check how sensitive the UFD bound is to the shape of PT, $\mathcal{P}_\zeta$, which differs from a delta function profile. For this, we first estimate the fraction of DM that has collapsed into clumps with masses within a decade of $M$ as
\begin{equation}
f_{\rm clump}(z,M)\approx M^2\frac{{\rm d}n}{{\rm d}M}\,,
\end{equation}
where we calculate the halo mass function using Eqs.~(\ref{eq.dndM}) and (\ref{eq.sig2}). Assuming that the DM clumps have  cores following the Navarro–Frenk–White (NFW) profile, Ref.~\cite{Graham:2024hah} estimates the density within the clump radius $R(M)$ as 
\begin{equation}
\rho_s(z)\approx\frac{\Delta}{2.317}\rho_m(1+z)^3\,.
\end{equation}
By comparing the PT prediction for $f_{\rm clump}(z,M)$ with the corresponding constraints on $f_{\rm clump}$ in Fig.~5 of~\cite{Graham:2024hah}, evaluated at given clump mass $M$ and density $\rho_s$, we obtain the solid red bound from the UFD heating in Fig.~\ref{fig:combined_bounds}\footnote{Instead of scanning the constraints over a continuous range of $\rho_s$, as done in Ref.~\cite{Graham:2024hah}, we evaluate the bounds only at the discrete $\rho_s$ values shown in their Fig.~5. As a cross-check, we reproduce the primordial power spectrum bound in their Fig.~6 assuming a delta-function–like shape, finding agreement at the $\sim20\%$ level.}. The UFD constraint extends the coverage of PT scenarios up to $T_{\rm PT}\simeq10^{5}$~eV for $\beta/H_{\rm PT} =10$.

\section{Conclusion}\label{sec:conclusion}
First-order phase transitions occurring at temperatures below $\mathcal{O}(100)$~keV can leave observable imprints on the CMB and the large- and small-scale structures. 
In this work, we derive and strengthen constraints on PTs occurring in a secluded dark sector, focusing on how stochastic fluctuations in the PT completion time, arising from bubble nucleation and expansion, source curvature perturbations.

Building upon earlier studies, we significantly strengthen these constraints by (i) performing a dedicated MCMC analysis of combined CMB and BAO data, (ii) incorporating Lyman-$\alpha$ forest bounds using compressed eBOSS data, and (iii) extending the sensitivity to higher PT temperatures through curvature perturbation constraints from early reionization and the observation of ultra-faint dwarf galaxies. By exploiting the imprints of phase transitions on cosmological perturbations, we derive substantially stronger bounds on PTs with $\beta/H_{\rm PT} =10$ compared to the homogeneous $\Delta N_{\rm eff}$ constraint, as well as provide bounds that are complementary to searches for $B$-modes induced by the gravitational wave background from a PT \cite{Greene:2024xgq, Zebrowski:2026pye}. Some of these perturbation-based bounds typically weaken and eventually become less restrictive compared to the $\Delta N_{\rm eff}$ limit for sufficiently large $\beta/H_{\rm PT}$, but CMB anisotropy measurements continue to provide a useful constraint even at $\beta/H_{\rm PT} =100$. Together, our constraints cover PTs with $\beta/H \lesssim 100$ over temperature ranges of $\sim1$ eV to $\sim10^5$ eV.

More broadly, our results demonstrate that any future improvement in probes of primordial curvature perturbations can open new avenues to test FOPT. A consistent detection of the predicted signatures—such as deviations in future Lyman-$\alpha$ measurements, exotic early reionization signals, or larger-than-expected half-light radii in UFD galaxies-would provide a compelling opportunity to probe curvature perturbations sourced by a post-inflationary process like the FOPT in the late Universe. In particular, next-generation CMB experiments such as the Simons Observatory~\cite{SO:2019} will dramatically improve sensitivity to small-scale perturbations in the CMB, while DESI~\cite{desiED_cat:2023,DESI_lyalpha_3D:2025,DESI_lyalpha_1DFFT:2025,DESI_lyalpha_1dOEM:2025,DESI_lyalpha_dr1valid:2026} and future Lyman-$\alpha$ surveys~\cite{NewHorizon:2025} will extend the reach to higher $k$-modes. Additionally, several proposed spectral distortion missions~\cite{Coulon:2024} could improve on FIRAS limits by several orders of magnitude, making $\mu$-distortions a powerful probe of late-time PTs in the dark sector.

A curvature perturbation $\zeta_{\rm PT}$ generated by a dark FOPT has several characteristic features. Unlike the almost scale-invariant spectrum predicted by standard inflationary models, the power spectrum for $\zeta_{\rm PT}$ exhibits a universal $k^3$ IR scaling on super-bubble scales--a direct consequence of causality that is largely insensitive to the microphysics of the transition. 
If future observations detect an excess of primordial spectrum, a scale dependence consistent with $k^3$ would be suggestive of a causal, potentially post-inflationary source. A detection of different IR scaling, such as the $k^1$ or $k^7$ slopes expected for matter- or kination-like dark fluids (Section~\ref{sec:darkFOPT}), could further reveal the equation of state of the dark sector populated by the FOPT. In addition, the presence of a finite bubble scale $d_b$ implies that $\zeta_{\rm PT}$ generated by stochastic bubble nucleation is inherently non-Gaussian, which might be probed in different cosmological surveys. Together, the characteristic spectral shapes of the primordial power spectrum and bispectrum offer a potential smoking-gun signature of a post-inflationary FOPT. A more detailed investigation of these signatures is left for future work.

\section*{Acknowledgment}
We thank Itamar Allali, Stuti Garg, Subhajit Ghosh, Peter Graham, Harikrishnan Ramani for the helpful discussions.  DH and YT are supported by the NSF Grant Number PHY-2412701. This research was supported in part by the Notre Dame Center for Research Computing. YT would also like to thank the Aspen Center for Physics (supported by NSF grant PHY-2210452). KG is supported by the Grant Korea RS-2024-00342093. This research was supported by the Munich Institute for Astro-, Particle and BioPhysics (MIAPbP) which is funded by the Deutsche Forschungsgemeinschaft (DFG, German Research Foundation) under Germany's Excellence Strategy – EXC-2094 – 390783311.

\appendix
\section{Calculation of curvature perturbation}
\label{app:Pnad}
Here we give more details in the derivation of Eq.~(\ref{eq:zeta_dot}). To describe the physics of this process on large scales, we need to express the perturbations in a gauge-invariant way that is independent of how one chooses to foliate space-time into certain hypersurfaces. A convenient quantity for this purpose is the curvature perturbation on uniform-density hypersurfaces: 
\begin{align}\label{eq:app_zeta_dot}
    \zeta(\vec{x}) \equiv -H\xi(\vec{x}) = -H\left(\frac{\psi(\vec{x})}{H}+\frac{\delta\rho(\vec{x})}{\dot{\rho}}\right)\,,
\end{align}
where $\xi$ is the gauge-invariant displacement between the uniform density ($\delta\rho = 0$) and spatially flat ($\psi=0$) hypersurfaces. In the presence of isocurvature perturbations, the curvature perturbation evolves on super-horizon scales as~\cite{Wands:2000dp}:
\begin{align}
    \dot{\zeta}(\vec{x}) = - \frac{H}{\rho + p}\delta p_{nad}(\vec{x})\,,
\end{align}
where $\delta p_{nad}$ is the non-adiabatic part of the total pressure perturbation 
\begin{align}
    \delta p_{nad}(\vec{x}) \equiv \dot{p}\,\Gamma(\vec{x}) = \dot{p}\left( \frac{\delta p(\vec{x})}{\dot{p}} - \frac{\delta\rho(\vec{x})}{\dot{\rho}}\right)
\end{align}
with $\Gamma$ the entropy perturbation. $\delta p_{nad}$ is non-vanishing if the equation of state $\omega \equiv p/\rho$ of the cosmic fluid can change independently of its density $\rho$. To see this, suppose that the pressure of the cosmic fluid goes as $p(\rho,s) = \omega(\rho,s)\rho$, where the equation of state has an additional dependence on some variable $s$ that is independent of $\rho$. Then
\begin{equation*}
    \delta p = \left(\omega + \frac{\partial\omega}{\partial\rho}\rho\right)\delta\rho  +\frac{\partial \omega}{\partial s}\rho \delta s =c_s^2\delta\rho + \delta p_{nad}
\end{equation*}
where the adiabatic sound speed is 
\begin{equation*}
    c_s^2 \equiv \frac{d p}{d \rho} = \omega + \frac{\partial\omega}{\partial\rho}\rho 
\end{equation*}
and $\delta p_{nad}$ corresponds to the remaining variation with respect to $s$. An FOPT introduces such perturbations since $F$ and $X$ redshift differently and the random transition time $t_c(\vec{x})$ provides the additional variable $s$ that $\omega$ can depend on. We may thus expect the variation $\delta p_{nad}(\vec{x})$ to go as $\delta t_c(\vec{x})$ across space.    

Performing the calculation in the $\beta \gg H_{\rm PT}$ regime, let us assume that the PT occurs almost instantaneously with the false vacuum $F$ being entirely converted into the dark fluid $X$ at $t_c(\vec{x})$. Denoting $a_c(\vec{x}) \equiv a(t_c(\vec{x}))$ and $\bar{a}_c \equiv a(\bar{t}_c)$ for the background scale factor, the energy density of the dark sector goes as
\begin{align}
    \rho_X(\vec{x}) = 
    \begin{cases}
        \rho_F \hspace{40mm}(t<t_c(\vec{x}))\\
        \rho_F\left( \frac{a(t)}{a_c(\vec{x})} \right)^{-(4+d_X)} \hspace{14mm} (t\geq t_c(\vec{x}))
    \end{cases}
\end{align}
with continuity at $t_c(\vec{x})$, while the standard model radiation bath goes as $\rho_{\rm SM} = \rho_{\rm SM, PT}(a/a_{\rm PT})^{-4}$ with reference scale $a_{\rm PT} = a(t_f)$. The total energy density and pressure in the radiation era are given by: 
\begin{align}
    \rho(\vec{x}) =& \rho_F \Theta(t_c(\vec{x}) - t) + \rho_{F}\left( \frac{a(t)}{a_c(\vec{x})} \right)^{-(4+d_X)} \Theta(t - t_c(\vec{x})) + \rho_{\rm SM, PT} \left( \frac{a(t)}{a_{\rm PT}} \right)^{-4}\,,\\
    p(\vec{x})
    =& -\rho_F \Theta(t_c(\vec{x}) - t) + \left(\frac{1}{3} + \frac{d_X}{3}\right)\rho_{F}\left( \frac{a(t)}{a_c(\vec{x})} \right)^{-(4+d_X)} \Theta(t - t_c(\vec{x})) + \frac{1}{3}\rho_{\rm SM, PT} \left( \frac{a(t)}{a_{\rm PT}} \right)^{-4}\nonumber\,,
\end{align}
where $\omega_{\rm SM} = 1/3$ for radiation, $\omega_F = -1$ for cosmological constant and $\omega_X = (1+d_X)/3$ for the dark fluid $X$. Notably, the abrupt change in the equation of state from $\omega_F$ to $\omega_X$ makes the pressure of the dark sector discontinuous at $t_c(\vec{x})$, which might be anticipated for a FOPT since pressure is a first-derivative of a thermodynamic potential.

As explained in the main text, the density and pressure perturbations can be obtained by varying these expressions with respect to $t_c(\vec{x})$ to linear order:
\begin{align}
    \delta\rho(\vec{x}) =&\ \rho_F \delta(\bar{t}_c - t)\delta t_c(\vec{x}) - \rho_{F}\left( \frac{a(t)}{\bar{a}_c} \right)^{-(4+d_X)}\delta(t - \bar{t}_c)\delta t_c(\vec{x})\\ &+ (4+d_X)\rho_{F}\left( \frac{a(t)}{\bar{a}_c} \right)^{-(4+d_X)} \Theta(t - \bar{t}_c)\frac{\delta a_c(\vec{x})}{\bar{a}_c}\,,\nonumber\\
    =&\ (4+d_X)\rho_{F}\left( \frac{a(t)}{\bar{a}_c} \right)^{-(4+d_X)} \Theta(t - \bar{t}_c)\frac{\delta t_c(\vec{x})}{2\bar{t}_c}\,,\nonumber\\
    \delta p(\vec{x}) =& -\rho_{F}\delta(\bar{t}_c-t)\delta t_c(\vec{x}) - \left(\frac{1}{3} + \frac{d_X}{3}\right)\rho_{F}\left( \frac{a(t)}{\bar{a}_c} \right)^{-(4+d_X)}\delta(t -\bar{t}_c)\delta t_c(\vec{x})\\
    &+ (4+d_X)\left(\frac{1}{3} + \frac{d_X}{3}\right)\rho_{F}\left( \frac{a(t)}{\bar{a}_c} \right)^{-(4+d_X)} \Theta(t - \bar{t}_c)\frac{\delta a_c(\vec{x})}{\bar{a}_c}\,,\nonumber\\
    =& -\left(\frac{4+d_X}{3}\right)\rho_{F}\delta(t -\bar{t}_c)\delta t_c(\vec{x}) + (1+d_X)\left(\frac{4+ d_X}{3}\right)\rho_{F}\left( \frac{a(t)}{\bar{a}_c} \right)^{-(4+d_X)} \Theta(t - \bar{t}_c)\frac{\delta t_c(\vec{x})}{2\bar{t}_c}\,,\nonumber
\end{align}
where we set $t=\bar{t}_c$ in terms multiplying the delta function $\delta(t -\bar{t}_c)$ and used $a\sim t^2$ in the radiation era. For the background quantities, we have 
\begin{align}
    \rho + p = \left(\frac{4+d_X}{3}\right)\rho_{F}\left( \frac{a(t)}{\bar{a}_c} \right)^{-(4+d_X)} \Theta(t - \bar{t}_c) + \frac{4}{3}\rho_{\rm SM, PT} \left( \frac{a(t)}{a_{\rm PT}} \right)^{-4}\,,
\end{align}
and by taking derivatives with respect to $t$:
\begin{align}
    \dot{\rho} 
    =&- (4 + d_X)\rho_{F}H\left( \frac{a(t)}{\bar{a}_c} \right)^{-(4+d_X)} \Theta(t - \bar{t}_c) - 4\rho_{\rm SM, PT} H\left( \frac{a(t)}{a_{\rm PT}} \right)^{-4}\,,\\
    \dot{p} =&\left(\frac{4+d_X}{3}\right)\rho_{F}\delta(t - \bar{t}_c)- (1 + d_X)\left(\frac{4+d_X}{3}\right)\rho_{F}H\left( \frac{a(t)}{\bar{a}_c} \right)^{-(4+d_X)} \Theta(t - \bar{t}_c) - \frac{4}{3}\rho_{\rm SM, PT} H\left( \frac{a(t)}{a_{\rm PT}} \right)^{-4}\,.\nonumber
\end{align}
One might notice in the calculations that only the variations in $p$ result in $\delta(t - \bar{t}_c)$ source terms while those in $\rho$ cancel exactly, which could have been anticipated from the fact that $\rho$ is continuous across the FOPT but $p$ is not. To simplify the subsequent calculations, we set $\bar{t}_c\approx t_f$ and $a_c \approx a_{\rm PT}$ to obtain
\begin{align}
    \delta\rho(\vec{x}) =&\ (4+d_X)\rho_{F}\left\{\left( \frac{a(t)}{a_{\rm PT}} \right)^{-(4+d_X)} \Theta(t - \bar{t}_c)\right\}\frac{\delta t_c(\vec{x})}{2\bar{t}_c}\,,\\
    \delta p(\vec{x}) =&\ \left(\frac{4+d_X}{3}\right)\rho_{F}\left\{ -\frac{1}{H_{\rm PT}}\delta(t -\bar{t}_c) + (1+d_X)\left( \frac{a(t)}{a_{\rm PT}} \right)^{-(4+d_X)} \Theta(t - \bar{t}_c)\right\} \frac{\delta t_c(\vec{x})}{2\bar{t}_c}\,,\\
     \dot{\rho} =& - 4\rho_{\rm SM, PT} H\left( \frac{a(t)}{a_{\rm PT}} \right)^{-4}\left\{1+ \frac{4 + d_X}{4}\alpha_{\rm PT}\left( \frac{a(t)}{a_{\rm PT}} \right)^{-d_X} \Theta(t - \bar{t}_c)\right\}\,.
\end{align}
\begin{equation}
    \hspace*{-15mm}\dot{p} = - \frac{4}{3}\rho_{\rm SM, PT} H\left( \frac{a(t)}{a_{\rm PT}} \right)^{-4}\left\{1 -\frac{4 + d_X}{4}\frac{\alpha_{\rm PT}}{H_{\rm PT}}\delta(t-\bar{t}_c) + (1+d_X)\left(\frac{4 + d_X}{4}\right)\alpha_{\rm PT}\left( \frac{a(t)}{a_{\rm PT}} \right)^{-d_X} \Theta(t - \bar{t}_c)\right\}\,. 
\end{equation}
Using these, we calculate the non-adiabatic pressure perturbation
\begin{align}
\hspace{-4cm}\delta p_{\rm nad}(\vec{x})
&= \delta p(\vec{x}) - \frac{\dot{p}}{\dot{\rho}}\,\delta\rho(\vec{x})\notag\\
&= \left(\frac{4+d_X}{3}\right)\rho_F
\Bigg\{
-\frac{1}{H_{\rm PT}}
\left[1 - 
\frac{\frac{4+d_X}{4}\,\alpha_{\rm PT}
\Theta(t-\bar{t}_c)}{
1 + \frac{4+d_X}{4}\,\alpha_{\rm PT}
\Theta(t-\bar{t}_c)
}
\right]
\delta(t-\bar{t}_c)\,,
\notag\\
&\hspace{1.5cm}
+ \left[1+d_X - 
\frac{
1 + (1+d_X)\frac{4+d_X}{4}\,\alpha_{\rm PT}
\left(\frac{a(t)}{a_{\rm PT}}\right)^{-d_X}
\Theta(t-\bar{t}_c)}{
1 + \frac{4+d_X}{4}\,\alpha_{\rm PT}
\left(\frac{a(t)}{a_{\rm PT}}\right)^{-d_X}
\Theta(t-\bar{t}_c)
}
\right]
\left(\frac{a(t)}{a_{\rm PT}}\right)^{-(4+d_X)}
\Theta(t-\bar{t}_c)
\Bigg\}
\frac{\delta t_c(\vec{x})}{2\bar{t}_c}\,.
\end{align}
where we set $a(t) = \bar{a}_c \approx a_{\rm PT}$ under the delta function. Canceling terms in the square brackets lead to:  
\begin{align}
\hspace{-4cm}\delta p_{\rm nad}(\vec{x})
&= \delta p(\vec{x}) - \frac{\dot{p}}{\dot{\rho}}\,\delta\rho(\vec{x})\notag\\
&= \left(\frac{4+d_X}{3}\right)\rho_F
\Bigg\{
-\frac{1}{H_{\rm PT}}
\left[
\frac{1}{
1 + \frac{4+d_X}{4}\,\alpha_{\rm PT}
\left(\frac{a(t)}{a_{\rm PT}}\right)^{-d_X}
\Theta(t-\bar{t}_c)
}
\right]
\delta(t-\bar{t}_c)\,,
\notag\\
&\hspace{3.5cm}
+ \left[
\frac{d_X}{
1 + \frac{4+d_X}{4}\,\alpha_{\rm PT}
\left(\frac{a(t)}{a_{\rm PT}}\right)^{-d_X}
\Theta(t-\bar{t}_c)
}
\right]
\left(\frac{a(t)}{a_{\rm PT}}\right)^{-(4+d_X)}
\Theta(t-\bar{t}_c)
\Bigg\}
\frac{\delta t_c(\vec{x})}{2\bar{t}_c}\,.
\end{align}
Plugging this back into Eq.~\eqref{eq:app_zeta_dot} together with
\begin{equation}
\rho + p =\ \frac{4}{3}\rho_{\rm SM, PT} \left( \frac{a(t)}{a_{\rm PT}} \right)^{-4}\left\{1 + \frac{4+d_X}{4}\alpha_{\rm PT}\left( \frac{a(t)}{a_{\rm PT}} \right)^{-d_X} \Theta(t - \bar{t}_c)\right\}\,,
\end{equation}
gives the source term of Eq.~(\ref{eq:zeta_dot}).

\section{Numerical calculation of the PT spectrum }
\label{app:numerical}
\begin{figure}
    \centering
    \includegraphics[width=0.9\linewidth]{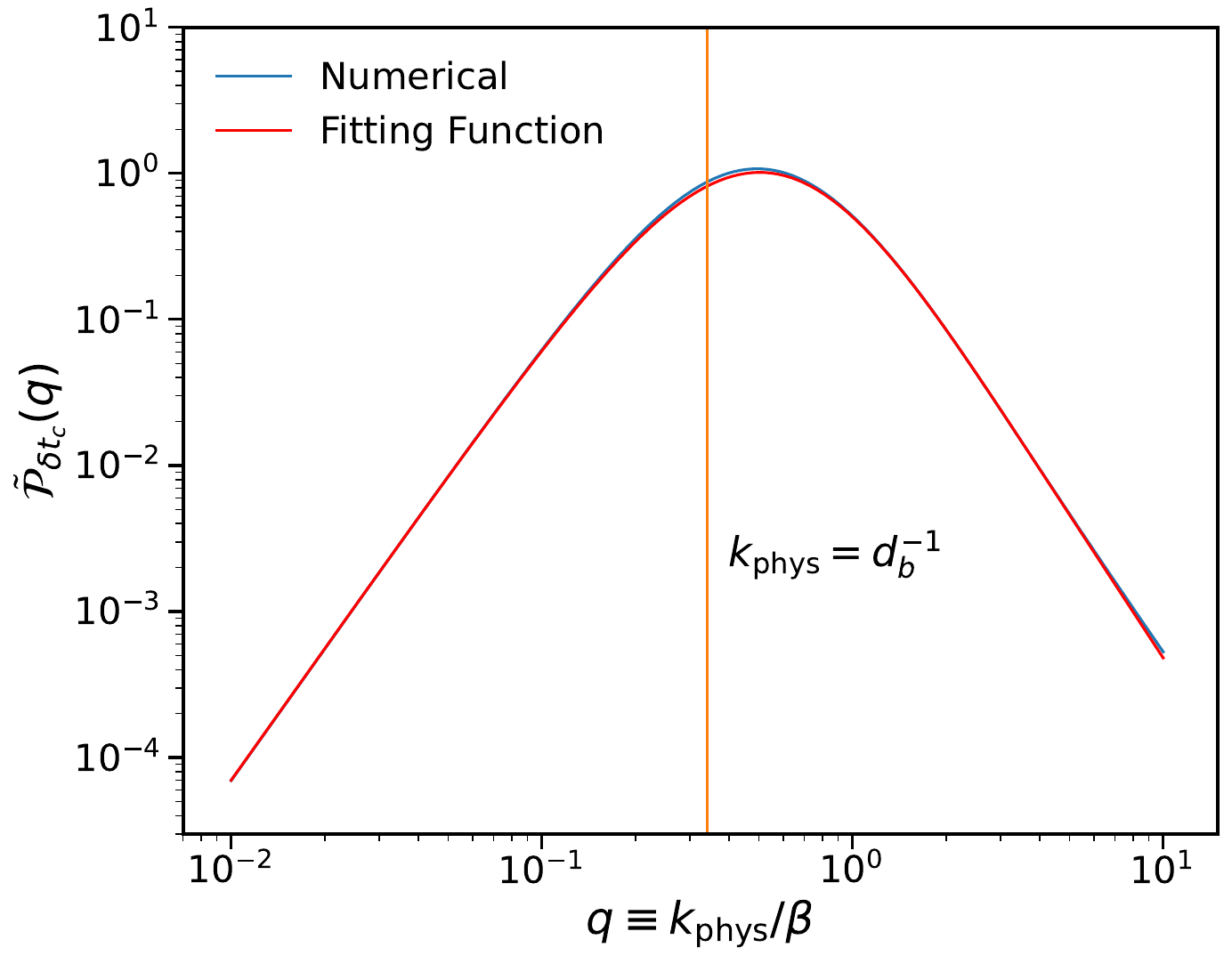}%\includegraphics[width=0.55\linewidth]{Figs/PkFit_Deviate.jpg}
    \caption{Plot of $\Tilde{{\cal P}}_{\delta t}(q)$ in the dimensionless variable $q\equiv k_{\rm phys}/\beta$ where $k_{\rm phys} = k/a_{\rm PT}$ is the physical wavenumber at PT. When implementing the PT signal in the \CLASS\ \texttt{external\_Pk} module, a fitting function to the spectrum was used with sub-percent deviation from the numerical result in the peak and super-bubble slope regions.
    % \KG{This is pretty tough to read as is for size. Could we make it all larger in accordance to my suggestions on figure 1, and move the equation out of the legend?}
    }
    \label{fig:numeric}
\end{figure}

We give more details on the numerical calculation of the dimensionless $\mathcal{P}_{\delta t}(k)$ spectrum in Eq.~\eqref{eq.Pdt}. Recall that $\mathcal{P}_{\delta t}(k)$ is the power spectrum corresponding to the two-point correlator $\hpt^2 \langle \delta t_c(\vec{x}) \delta t_c(\vec{y}) \rangle$ for PT time fluctuations. It is convenient to separate out the background expansion to isolate the FOPT bubble dynamics $\mathcal{P}_{\delta t}(k) = \left({\hpt}/{\beta}\right)^2 \Tilde{\mathcal{P}}_{\delta t}(k)$, where $\Tilde{\mathcal{P}}_{\delta t}(k)$ is the power spectrum for the dimensionless quantity $\beta\delta t_c$: 
\begin{equation}
\Tilde{\mathcal{P}}_{\delta t}(k) = \frac{k^3}{2\pi^2} \int \mathrm{d}^3 r~ e^{i \vec{k} \cdot \vec{r}} \beta^2 \langle \delta t_c(\vec{x}) \delta t_c(\vec{y}) \rangle 
\end{equation}
To calculate $\Tilde{\mathcal{P}}_{\delta t}(k)$ numerically, we rescale the coordinates by $\beta$ to work in dimensionless variables $v \equiv \beta t$, $u \equiv \beta r_{\rm phys}$ and $q \equiv k_{\rm phys}/\beta$, where $r_{\rm phys} = r a_{\rm PT}$ and $k_{\rm phys} = k/a_{\rm PT}$ are the physical displacement and physical wavenumber at PT respectively. Assuming statistical homogeneity and isotropy:
\begin{equation}\label{eq:tilde_P}
\Tilde{\mathcal{P}}_{\delta t}(q) = \frac{q^3}{2\pi^2} \int_0^\infty \mathrm{d}u\ 4\pi u^2\ \frac{\sin (qu)}{qu} \langle\beta\delta t_c\beta\delta t_c\rangle(u)
\end{equation}
The explicit form of $\langle\beta\delta t_c\beta\delta t_c\rangle(u)$ is derived in the supplemental materials of Ref.~\cite{Elor:2023xbz} by analyzing the light cone geometry of the nucleated bubbles. We quote the final results here:
\begin{equation}\label{eq:tilde_dt}
    \langle\beta\delta t_c\beta\delta t_c\rangle(u) = \int^{u}_{-u}dv\big( A(u,v) +  B(u,v)\big)
\end{equation}
where 
\begin{align*}
\hspace{-2cm}A(u,v) &=  \frac{2\pi e^{-u/2}}{uI(u,v)}\left(\frac{u^2}{4}+u+2-\frac{v^2}{4}\right)\left[\left(\frac{\ln(I(u,v))}{8\pi}\right)^2-\frac{v^2}{4}+\frac{\pi^2}{6}\right]
\end{align*}
\begin{align*}
\hspace{-0.5cm}B(u,v) &= \frac{16\pi^2}{I(u,v)^2}\Bigg(4-\frac{e^{-u/2-v/2}}{2u}(u+v+4)(u-v)-\frac{e^{-u/2+v/2}}{2u}(u-v+4)(u+v)\notag\\
&\hspace{0.5cm}+\frac{e^{-u}}{16u^2}((u+4)^2-v^2)(u^2-v^2)\Bigg)\left[\left(\ln\left(\frac{I(u,v)}{(8\pi)}\right)-1\right)^2-\frac{v^2}{4}+\frac{\pi^2}{6}-1\right]
\end{align*}
and
\begin{align*}
I(u,v) = 8\pi\left(e^{v/2} + e^{-v/2} + \frac{v^2-(u^2+4u)}{4u}e^{-u/2}\right)
\end{align*}
$\Tilde{\mathcal{P}}_{\delta t}(q)$ is calculated by numerically integrating over $v$ in Eq.~\ref{eq:tilde_dt} and then over $u$ in Eq.~\ref{eq:tilde_P} for a range of $q$. The result in dimensionless $q$-space is presented in Fig~\ref{fig:numeric}. The full ${\cal P}_{\zeta, \rm PT}(k)$ in comoving wavenumber $k$ is then obtained by scaling this result with respect to the PT parameters. When implementing the spectrum in the \texttt{external\_Pk} module for \CLASS\ calculations, a fitting function for $\Tilde{\mathcal{P}}_{\delta t}(q)$ was used to allow for efficient and stable rescaling under parameter scans. This fitting function was obtained within sub-percent deviations from the numerical result in the peak and super-bubble slope regions.   

%%%%%%%%%%%%%%%%%%%%%%%%%%%%%%
\bibliography{bib-ref}
\bibliographystyle{JHEP}
%%%%%%%%%%%%%%%%%%
\end{document}